\documentclass[12pt,epsf]{article}

\usepackage{enumitem}   
\usepackage{graphicx,graphics,cite}
\usepackage{subfigure}
\newcommand{\ba}{\begin{array}}
\newcommand{\ea}{\end{array}}

\def\lQ{\Lambda_{\rm QCD}}

\newcommand{\be}{\begin{equation}}
\newcommand{\ee}{\end{equation}}
\newcommand{\bea}{\begin{eqnarray}}
\newcommand{\eea}{\end{eqnarray}}
\def\al{\alpha}
\def\als{\alpha_{\rm s}}
\def\siml{{\ \lower-1.2pt\vbox{\hbox{\rlap{$<$}\lower6pt\vbox{\hbox{$\sim$}}}}\ }} 
\def\simg{{\ \lower-1.2pt\vbox{\hbox{\rlap{$>$}\lower6pt\vbox{\hbox{$\sim$}}}}\ }} 

\newcommand{\MS}{\overline{\rm MS}}
\newcommand{\RS}{\rm RS}
\newcommand{\PV}{\rm PV}

\begin{document}
\title{\vskip-3cm{\baselineskip14pt
}
\vskip1.5cm
Bjorken sum rule with hyperasymptotic precision}
\author{Cesar Ayala$^{1}$ and 
Antonio Pineda$^{2,3}$\\[0.5cm]
{\small ${}^1$  \it Instituto de Alta Investigaci\'on, Sede Esmeralda, Universidad de Tarapac\'a,}\\
{\small \it Av. Luis Emilio Recabarren 2477, Iquique, Chile} \vspace{0.3cm}\\
{\small ${}^2$ \it Institut de Física d'Altes Energies (IFAE),}
{\small\it The Barcelona Institute of Science and Technology,}\\
{\small\it Campus UAB, 08193 Bellaterra (Barcelona), Spain}\\
{\small ${}^3$ \it Grup de Física Teòrica, Dept. Física,}\\
{\small \it Universitat Autònoma de Barcelona, E-08193 Bellaterra, Barcelona, Spain}}
\date{}

\maketitle

\thispagestyle{empty}

\begin{abstract}
We obtain an improved determination of the normalization of the leading infrared renormalon of the Bjorken sum rule: $Z_B (n_f=3)=  -0.407\pm 0.119 $. Estimates of higher order terms of the perturbative series are given. We compute the 
Bjorken sum rule with hyperasymptotic precision by including the leading terminant, associated with the first infrared renormalon. We fit the experimental data to the operator product expansion theoretical prediction with 
$\hat f_{3}^{\PV}$ as the free parameter. We obtain a good agreement with the experiment with 
$\hat f_{3}^{\PV}\times 10^3=32^{+187}_{-196}\;{\rm GeV}^{2}$ for $Q^2 \geq 1$ GeV$^2$.
\\[2mm]
\end{abstract}

\newpage

\section{Introduction}
\label{int}

The Bjorken sum rule \cite{Bjorken} is one of the cleanest observables, from the theory point of view, on which to apply the operator product expansion (OPE) in its nonperturbative version \cite{Shifman:1978bx}. At low orders in the $1/Q^2$ expansion, it has the following form:
\be
\label{Bjorken}
M_1^B(Q^2)={g_A \over 6}C_B(\als)
-{4 \over 27}{1 \over Q^2}f_3(Q_0)
\left[{\als(Q_0^2)\over \als(Q^2)}\right]^{-\gamma_{\rm NS}^0 \over 2\beta_0}
\left(1+{\cal O}(\als)\right)
+{\cal O}\left({1 \over Q^4}\right)
\,,
\ee
where the definitions are the following:\footnote{We will borrow the notation we use in this paper for the Bjorken sum rule from \cite{Campanario:2005np}. See also this reference for extra details on the sum rules.}
\be
\gamma_{NS}^0={16 \over 3}C_F,
\ee
\be
C_B(Q)=1+\sum_{s=0}^{\infty}C_B^{(s)}\als^{s+1}(\nu)\,.
\label{pertubativeseries}
\ee
The first few terms of the perturbative expansion of $C_B(Q)$ are known. The ${\cal O}(\alpha_s)$ correction has been computed in ~\cite{bj1loop},
the ${\cal O}(\alpha_s^2)$ correction in~\cite{bj2loop}, the ${\cal O}(\alpha_s^3)$ correction~\cite{bj3loop}, and the  ${\cal O}(\alpha_s^4)$ correction in ~\cite{Baikov:2010je,Larin:2013yba,Baikov:2010iw}.

We also consider the correction due to the charm quark (with finite mass) to the 
perturbative series of $C_B$. It was first computed in Ref. 
\cite{Blumlein:1998sh}, and later corrected in Ref. \cite{Blumlein:2016xcy}. Arguments from renormalons in the context of heavy quark physics suggest that the charm decouples at those scales for the high order coefficients of the perturbative expansion\cite{Ayala:2014yxa}. Therefore, this is the situation we will consider in this paper. In any case, the effect of these corrections is tiny. 

Power corrections were calculated in~\cite{highertwists,powercorrections}.
The LO renormalization group running of the twist-four operators has been computed 
in Ref.~\cite{Kawamura:1996gg}. The nonperturbative matrix elements are defined 
in the following way:
\begin{equation}
\label{gaa8}
 \begin{array}{llrll}
 |g_A| s_\sigma & = & 2 \langle p,s | J^{5,3}_\sigma | p,s \rangle  , \nonumber 
\end{array} 
\end{equation}
where $J_\sigma^{5,a}(x) = \overline{\psi} \gamma_\sigma
 \gamma_5 t^{a} \psi(x)$ is the nonsinglet axial current, $t^a$ is a
generator of the flavor group, and $J_\sigma^5(x) =
  \sum_{i=1}^{n_f} \overline{\psi_i} \gamma_\sigma \gamma_5 \psi_i(x)$
is the singlet axial current. $|g_A|$ is the absolute value of the constant of 
the neutron beta decay, $|g_A/g_V| = F+D= 1.2754  \pm  0.0013 $ \cite{ParticleDataGroup:2020ssz}. 

 The dimension two condensate $f_3$ can be related with the expectation value (sandwiched between the proton state) of a local operator. $f_3$ is scale dependent and it is defined at $Q_0^2$, 
i.e., $f_3$ is the reduced matrix element of $R_{2\sigma}^3$,
renormalized at $Q_0^2$, which is defined for the general flavor
indices, with $t^i$ being the flavor matrices, as 
\be
R_{2\sigma}^i=g\overline{\psi}\tilde{G}_{\sigma\nu}\gamma^{\nu}t^i\psi,
\quad \langle p,s|R_{2\sigma}^i|p,s \rangle=f_i s_{\sigma} 
\quad (i=0,\cdots,8),
\ee
and $\tilde{G}_{\mu\nu}=\frac{1}{2}\varepsilon_{\mu\nu\alpha\beta}G^{\alpha\beta}$
is the dual field strength.

Each term of the OPE expression written in Eq. (\ref{Bjorken}) is ill defined. Therefore, as such, the OPE is a formal expression, and, without further qualifications, it is of little use. The origin of this problem comes because, even though the Bjorken sum rule is an observable, and, therefore, is well defined, the splitting between the different terms of the OPE is ambiguous. When working in dimensional regularization using minimal-like subtraction schemes, this reflects in the fact that the perturbative series of $C_B(\als)$ is asymptotically divergent. Therefore, its sum does not converge to a number, and a method has to be used to regularize the perturbative sum. A suitable one is to first construct the Borel transform of the perturbative sum and, afterwards, to do the inverse of the Borel transform (also named Borel sum or integral). Such inverse is still ill defined, due to singularities in the positive axis of the Borel plane. The location and 
character of such singularities are determined by the OPE \cite{Parisi:1978bj,Mueller:1993pa}, but not the overall normalization. This information is enough to determine the divergence pattern of the perturbative series up to an overall normalization (for the case of the Bjorken sum rule, it has been computed with logarithmic accuracy in \cite{Campanario:2005np}). Such divergent pattern of the perturbative expansion associated with the OPE is usually referred to as renormalons \cite{tHooft}.

To fully fix the leading term of the OPE, we have to specify how we handle the singularities in the Borel plane when doing the Borel sum. Following the discussion in \cite{Ayala:2019uaw}, we use the principal value (PV) prescription for the Borel integral (the median resummation above and below the real axis). The reason is that the outcome is expected to be real and scale/scheme independent (see the discussion in Refs. \cite{Ayala:2019uaw,Takaura:2020byt}). 

Defining $C_B(\als)$ also defines $f_3$. Therefore, we rewrite Eq. (\ref{Bjorken}) as
\be
\label{BjorkenPV}
M_1^B(Q^2)={g_A \over 6}C_B^{\PV}(\als)
-{4 \over 27}{1 \over Q^2}{\hat f}_3^{\rm PV} \left[\als(Q^2)\right]^{\gamma_{\rm NS}^0 \over 2\beta_0}
\left(1+{\cal O}(\als)\right)
+{\cal O}\left({1 \over Q^4}\right)
\,,
\ee
where (often we will work with the variable $u=\frac{\beta_0 t}{4\pi}$ instead of $t$)
\be
\label{CBPV}
C_B^{\PV}(\als(Q))=1+\int_{0,\rm PV}^{\infty} dt\; e^{-t/\als(Q)}B[C_B](t)\,, \qquad B[C_B](t)=\sum_{n=0}^{\infty}
\frac{C_B^{(s)}}{s!}t^s
\,,
\ee
and
\be
{\hat f}_3^{\rm PV}\equiv f_3^{\rm PV}(Q_0)\left[\als(Q_0^2)\right]^{-\gamma_{\rm NS}^0 \over 2\beta_0}
\,.
\ee
For convenience, we have absorbed the prefactor $\left[\als(Q_0^2)\right]^{-\gamma_{\rm NS}^0 \over 2\beta_0}$ in  
$\hat f_3^{\rm PV}$. It is usually stated that, after introducing such prefactor, $\hat f_3$ is a renormalization group invariant (with the precision the running of the nonperturbative condensate is known). We would like to emphasize that this is not necessarily so. In order $\hat f_3$ to be renormalization scale and scheme independent, it is necessary that the regularization of the perturbative sum of $C_B$ is made in such a way that it is explicitly scale and scheme independent (something that the PV delivers but not other regularizations of the perturbative sum), so that the difference between the Bjorken sum rule and $\frac{g_A}{6}C_B(\als(Q))$ is also scale and scheme independent. 

Assuming the validity of the OPE in its nonperturbative version (which we take for granted), $\hat f_3^{\rm PV}$ can be written as $\hat f_3^{\rm PV}=\bar f_{3,\MS}^{\rm PV} \Lambda_{\MS}^2$, where $\bar f_{3,\MS}^{\rm PV}$ is a dimensionless constant. Let us emphasize that such equality holds irrespectively of the scheme used for the strong coupling constant. Therefore, we can also generically write 
\be
\label{eq:hatf3PV}
\hat f_3^{\rm PV}=\bar f_{3,X}^{\rm PV} \Lambda_{\rm X}^2=\bar f_{3,X}^{\rm PV}
\left(\frac{\beta_0\alpha_X(\nu)}{4\pi}\right)^{-2b}e^{-\frac{4\pi}{\beta_0\alpha_X(\nu)}}\left(1+{\cal O}(\alpha_X)\right)
\,,
\ee
 where X stands for the renormalization scheme of the strong coupling constant and 
we define $b={\beta_1/(2\beta_0^2)}$ (note that the definition of $b$ that we use in this paper is different from the one used in Ref. \cite{Campanario:2005np}). Therefore, provided $\bar f_{3,\rm X}^{\rm PV}$ is obtained in one scheme, one can easily transform it to a different scheme using the very same conversion factor one uses to transform $\lQ$ from one scheme to another. In the present work $X=\MS$.

Whereas the above procedure yields unambiguous and convenient definitions of the different terms of the OPE, this does not mean that we have the means to compute them. In practice, we only have a relatively small set of the first order terms of the perturbative expansion, and the experimental (or lattice) data. Nowadays, it is not possible to compute the nonperturbative corrections from first principles via analytic methods. At present, it is only possible to compute them in some cases, numerically, from lattice simulations. Nevertheless, such computation is unavoidably plagued by perturbative corrections. Actually, those are the dominant contribution to the observable. A paradigmatic case is the computation of the gluon condensate in the lattice, where its value is orders of magnitude bigger than the actual size of the nonperturbative gluon condensate \cite{Bali:2014sja,Ayala:2020pxq}. Overall, to fit the nonperturbative condensate to lattice or experimental data, one first needs to compute the perturbative series with exponential [in $-1/\al(Q)$] accuracy (or with power in $1/Q^2$ accuracy). Whereas it is not possible to obtain the exact expression of $C_B^{\rm PV}$, it can be computed approximately, and, more importantly, with a well-defined method to quantify the error in a parametric way \cite{Ayala:2019uaw,Ayala:2019lak,Ayala:2019hkn}. This method adapts the hyperasymptotic expansion used in ordinary differential equations \cite{BerryandHowls} (see also \cite{Dingle}) to the case of quantum field theories with marginal operators, and it has successfully been applied to a variety of observables \cite{Ayala:2019hkn,Ayala:2020odx,Ayala:2020pxq}. The hyperasymptotic approximation has not yet been confronted directly to experimental data. The Bjorken sum rule yields a fantastic opportunity to check the OPE in its nonperturbative setup directly into experiment, without resorting to lattice. Therefore, it is our plan to apply such a method to the Bjorken sum rule. Nowadays, the perturbative series of $C_B(\als(Q)$ is known to high enough orders to start showing its asymptotic nature for the range of energies for which such sum rule has been measured. Therefore, this opens the venue to determine the Borjen sum rule with exponential accuracy. This is very interesting, since it will allow us to determine the leading nonperturbative correction with hyperasymptotic precision. 

Overall, we have now turned the problem into evaluating $C_B^{\PV}$ with the highest possible accuracy, i.e., including the leading terminant. We address this goal in Sec. \ref{Sec:CBPV}. One necessary ingredient in the evaluation of the terminant is the determination of the normalization of the renormalon. A previous determination can be found in \cite{Campanario:2005np}. In the present paper, we give an improved determination. On the one hand, we can use the new term of the perturbative expansion that is now known. Another improvement is a new way to determine the normalization of the renormalon, which has proven to be more stable \cite{Bali:2013pla}. As this method had not been applied to the GLS \cite{GLS69} and Ellis-Jaffe \cite{EllJaf74} sum rule, we profit this paper to do the same analysis for those sum rules. For the case of the GLS sum rule, the perturbative expansion is also known to an order higher \cite{Baikov:2010iw}, but not for the Ellis-Jaffe sum rule. The latter has also been criticized as being less fundamental than the Bjorken and GLS sum rules (see \cite{Blumlein:2012bf}).

The paper is organized as follows. 
In Sec. II, $C_B^{\PV}$ will be computed with hyperasymptotic precision, and an updated determination of the normalization of the leading infrared renormalon, as well as new estimates of the higher order terms of the pertubative series, are given. Updated determinations of the normalization of the leading renormalons of the Ellis-Jaffe and GLS sum rules are also given.
In Sec. III, the comparison with the experimental data will be done allowing us the extraction of $\hat f_3^{\rm PV}$. Finally, 
the conclusions are presented in Sec. IV.

\section{Hyperasymptotic approximation to $C_B^{\PV}$}
\label{Sec:CBPV}

\subsection{Renormalons}
We first need to know the renormalon structure of the perturbative series. We take the results relevant to our case from the analysis made in Ref. \cite{Campanario:2005np}.

The ultraviolet renormalon structure of the moments of the deep inelastic scattering structure functions has 
been computed in Ref.~\cite{Beneke:1997qd}. For the case of the Bjorken sum rule, the ultraviolet renormalon 
formally dominates for $n_f > 2$ and $n \rightarrow \infty$. Nevertheless, 
at low orders in perturbation theory, 
the infrared renormalon appears to be dominant. This can be seen 
from the fact that the sign of the known terms of the perturbative series is 
equal, whereas if the ultraviolet renormalon were to be dominant, then we would find a sign alternating series. In any case, we will perform the conformal mapping~\cite{Contreras:2002kf}:
\be
\label{CMw}
w(u)=\frac{\sqrt{1+u}-\sqrt{1-u/2}}{\sqrt{1+u}+\sqrt{1-u/2}}
\,,
\ee
to 
 guarantee that the $u=-1$ renormalon does not play a significant role. This transformation maps the first infrared renormalon to $w=1/3$ and all other singularities to 
the unit circle $|w|=1$. In the conformal mapping, the expansion parameter is $w=1/3$.
In practice, the effect of doing the conformal mapping is small, which, again, points to the fact that 
the effect of ultraviolet renormalons is small in comparison with the effect of the
infrared renormalon located at $u=1$. A similar conclusion was obtained in Ref.~\cite{Ellis:1995jv} using 
Pade approximants. We will only give numbers for the conformal mapping case 
for theoretical reasons.  Nevertheless, as we have already mentioned, they will be quite similar to the 
computation without conformal mapping. 

The Borel transform near the closest infrared renormalon singularity has the following structure (we have changed $N_X \rightarrow Z_X$ compared with the notation in Ref. \cite{Campanario:2005np} to avoid confusion with $N_P$):
\begin{equation}
\label{SX}
B[C_B](t(u))= {\nu^2 \over Q^2}Z_B
\frac{1}{(a-u)^{1+2b+b_B}}(1+d_1^B(a-u)+d_2^B(a-u)^2+\cdots) +S_{reg}(u)
\,,
\end{equation}
where $S_{reg}(u)$ is an analytic function at $u=a$. $a=1$ and \cite{Mueller:1993pa}:
\be
b_{B}
=-{\gamma_{NS}^0 \over 2\beta_0}\,.
\ee
$b_B$ dictates the strength of the singularity. It is interesting 
to study its dependence on $n_f$. In the Bjorken sum rule, 
if $n_f \in (0,6) \Rightarrow 
1+2b+b_B \in (1,2)$ 
so, formally, one could just keep the first two terms of the series in 
Eq.~(\ref{SX}), since the next term would go 
to zero for $u \rightarrow 1$. 

If the Wilson coefficients multiplying the 
higher twist operators were known exactly, then one could also fix the 
coefficients $d_r^B$. Unfortunately, only their leading log running is known.
Nevertheless, by performing the matching at a generic scale $\nu$, one can resum the terms of the type $(1-u)^n\ln^n(Q^2/\nu^2)$ and 
obtain the logarithmically dominant contribution to $d_r^B \sim 
\ln^r(Q^2/\nu^2)$. This was done in \cite{Campanario:2005np}, and we have little to add in this respect in this paper. The outcome for the asymptotic expression of the coefficients was the following:
\be
\label{CXRG}
C_B^{(n)}
\stackrel{n\rightarrow\infty}{=} Z_B\,{\nu^2 \over Q^2}\,
\left({\beta_0 \over 4\pi}\right)^n
\,{\Gamma(1+b_B+2b+n) \over
\Gamma(1+2b+b_B)}
{}_1F_1\left(-b_B,-2b-b_B-n,\ln(Q^2/\nu^2)\right) 
.
\ee
The above expression contains subleading terms in the $1/n$ expansion. 
In the strict $1/n$ expansion, 
it simplifies to:
\be
\label{CXRG1overn}
C_B^{(n)}
\stackrel{n\rightarrow\infty}{=} Z_B\,{\nu^2 \over Q^2}\,
\left({\beta_0 \over 4\pi}\right)^n
\, {n! \,n^{b_B+2b} \over
\Gamma(1+2b+b_B)}
\left(1+{1 \over n}\ln(Q^2/\nu^2)\right)^{b_B} 
.
\ee

An alternative that we will consider in this paper is to work with the quantity 
\be
\label{tildeCB}
\tilde  C_B(\als) \equiv (C_B(\als)-1)\left[\frac{\als(Q)}{\als(\nu)}\right]^{b_B}=\sum_{n=0}^{\infty}\tilde C_B^{(n)}\als^{n+1}
\,,
\ee
the coefficients of which have a more compact expression for its asymptotic expansion:
\be
\label{tildeCXRG}
\tilde C_B^{(n)}
\stackrel{n\rightarrow\infty}{=} Z_B\,{\nu^2 \over Q^2}\,
\left({\beta_0 \over 4\pi}\right)^n
\,{\Gamma(1+b_B+2b+n) \over
\Gamma(1+2b+b_B)} 
\,,
\ee
or
\be
\label{tildeCXRG1overn}
\tilde C_B^{(n)}
\stackrel{n\rightarrow\infty}{=} Z_B\,{\nu^2 \over Q^2}\,
\left({\beta_0 \over 4\pi}\right)^n
\, {n! \,n^{b_B+2b} \over
\Gamma(1+2b+b_B)}
.
\ee

As a final comment, the whole previous discussion also applies to the GLS and Ellis-Jaffe sum rules changing any subscript $B$ by the $GLS$ and $EJ$ subscripts respectively. In particular, $b_{GLS}=b_B$ and 
$b_{EJ}=-\gamma_S^0/(2\beta_0)$ ($\gamma_S^0=\gamma_{NS}^0+4/3n_f$). Note also that for the case $n_f=0$, the perturbative expansions of the three sum rules coincide, as differences between singlet ($S$) and nonsinglet ($NS$) have to do with light fermion effects, which disappear in the $n_f=0$ approximation. 

\subsection{Determination of the normalization constant}

In this subsection, we will obtain the normalization of the leading infrared renormalon for the Borjken, GLS, and Ellis-Jaffe sum rules: $Z_B$, $Z_{GLS}$, and $Z_{EJ}$. For short, we will use the notation $Z_X$, where $X=\{B,GLS,EJ\}$.
A previous determination of these quantities was obtained in Ref. \cite{Campanario:2005np}. In such reference, 
the ultraviolet renormalon was neglected, and the normalization of the leading infrared renormalon was obtained using the following equality (actually, using the analogous expression after doing the conformal mapping of Eq. (\ref{CMw})):
\be
\label{DX}
D_X(u=1)=\sum_{n=0}^{\infty}D_X^{(n)}=Z_X{\nu^2 \over Q^2}
\,,
\ee
where
\be
D_X(u)=(1-u)^{1+2b+b_X}B[C_X](t(u))=\sum_{n=0}^{\infty}D_X^{(n)}u^n
\,.
\ee
It has been observed in Refs. \cite{Bali:2013pla,Ayala:2014yxa} that a more stable result is obtained by determining the normalization directly from the ratio of the exact and asymptotic expression. Therefore, this is the path we will follow in this paper instead. The results will not be very different. 
To eliminate the anomalous dimension of the leading infrared renormalon, we consider the perturbative series of $\tilde C_B$ (for consistency, the prefactor is expanded with one-loop accuracy), defined in Eq. (\ref{tildeCB}), and take the asymptotic expression given in Eq. (\ref{tildeCXRG}).
To both expressions we do the conformal mapping and consider the ratio of the corresponding coefficients at $x=1$ with NNNLO precision, where $x=\nu/Q$. These we will take as our central values. Actually, for the asymptotic expression, we already take the expected asymptotic expression after conformal mapping, i.e., with the leading singularity located at $w=1/3$ [this is nothing but multiplying Eq. (\ref{tildeCXRG}) by $3^n$]. We have checked that the difference is indeed small, and well inside of what would be our errors. We have also checked that not doing the conformal mapping does not change our determination significantly, which is consistent with the interpretation that the ultraviolet renormalon is not important. Another check of this interpretation comes from taking the $n_f \rightarrow \infty$ limit of our results. These should converge to the large $\beta_0$ results obtained in \cite{Broadhurst:1993ru}. Indeed, this is what approximately happens: the result converges to the result expected from the large $\beta_0$ analysis for the infrared renormalon. Actually, this also happens even if we do not do the conformal mapping. 

The error in our determination of the normalization of the infrared renormalon is due to the incomplete knowledge of the perturbative series. This reflects in that the result will depend on the scale, the order we truncate, and the explicit expression we use to determine the normalization. We use these three methods as indicators of the error. In particular, we do the following:
\begin{enumerate}[label=(\roman*)]
\item Scale variation between $x \in (1/2,2)$.

\item Difference between NNLO and NNNLO at $x=1$.

\item We consider the difference between using (the conformal transform of) Eqs. (\ref{tildeCXRG}) and (\ref{tildeCXRG1overn}) for the asymptotic coefficient (as they generate different $1/n$ corrections). 
\end{enumerate}
The three methods yield different ways to measure the fact that $n$ is not infinity. We take the largest of the three as our estimate of the error.  The scale variation will yield the largest error that we then symmetrize to its maximum value. 

\begin{table}[htb!]
\begin{center}
\begin{tabular}{|c|c|c|c|}
\hline
$n_f$&$Z_B$ &$Z_{GLS}$&$Z_{EJ}$\\\hline
0&-0.506$\pm$ 0.186&-0.506$\pm$0.186&-0.506$\pm$0.186\\ \hline
1&-0.480$\pm$ 0.169&-0.473$\pm$0.165&-0.346$\pm$0.171\\ \hline
2&-0.449$\pm$ 0.149&-0.432$\pm$0.138&-0.249$\pm$0.125\\ \hline
3&-0.407$\pm$ 0.119&-0.374$\pm$0.098&-0.099$\pm$0.047\\ \hline
4&-0.339$\pm$ 0.097&-0.279$\pm$0.158&0.176$\pm$0.105\\ \hline
5&-0.224$\pm$ 0.203&-0.113$\pm$0.323&0.816$\pm$0.485\\ \hline
6&0.037$\pm$ 0.441&0.256$\pm$0.700&3.18$\pm$2.80\\ \hline
\end{tabular}
\end{center}
\caption{\it Values of the infrared renormalon residue of the Bjorken, 
Ellis-Jaffe and 
GLS leading-twist Wilson 
coefficient $C_X$. For $n_f=0$ the three are equal. Note that the determination for the Ellis-Jaffe sum rule is obtained at one order less than the other normalizations (except for the case $n_f=0$).}
\label{tableNX}
\end{table}

\begin{figure}[htb!]
\vspace{-40pt}
\hspace{90pt}
\subfigure[{$Z_B$}, $Z_{EJ}$ and $Z_{GLS}$ with $n_f=0$.]{\includegraphics[scale
=0.75]{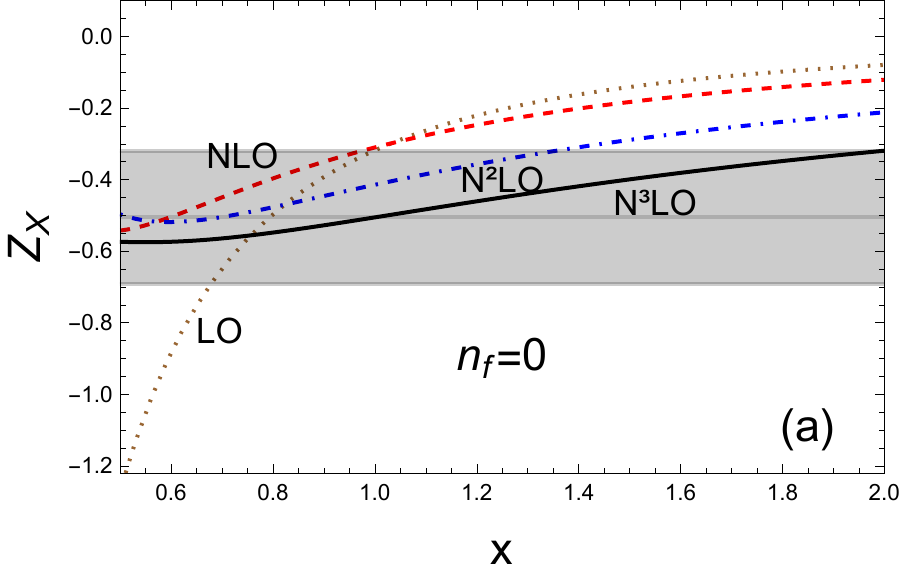}} 
\\
\begin{tabular}{cr}
\hspace{-20pt}
\subfigure[{$Z_B$} with $n_f=3$.]{\includegraphics[scale
=0.75]{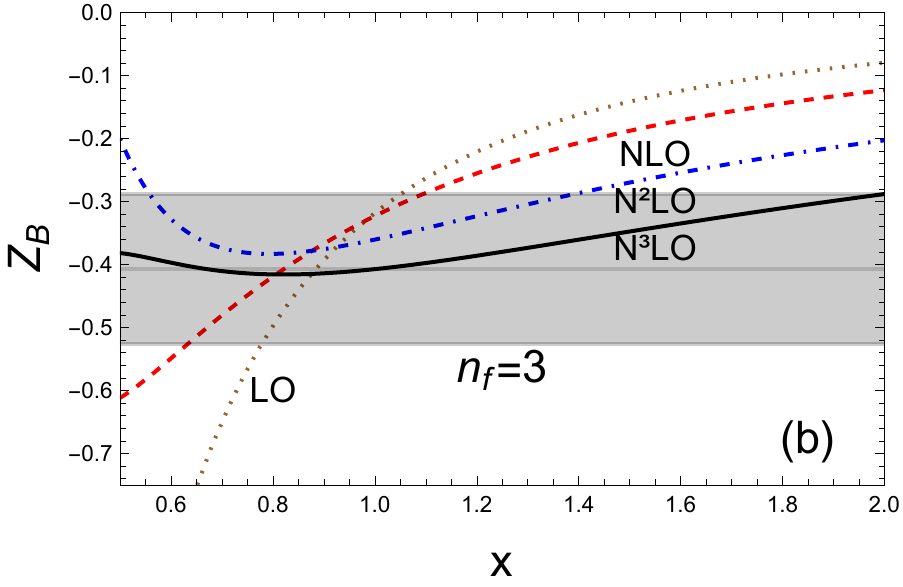}} &~
\hspace{25pt}
\subfigure[{$Z_B$} with $n_f=6$.]{\includegraphics[scale
=0.75]{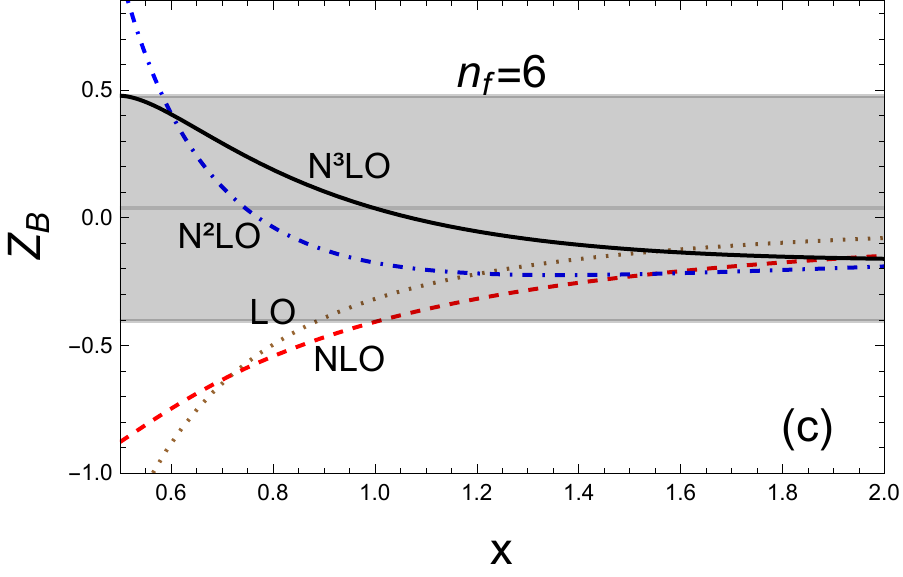}}
\\
\hspace{-20pt}
\subfigure[{$Z_GLS$} with $n_f=3$.]{\includegraphics[scale=0.75]{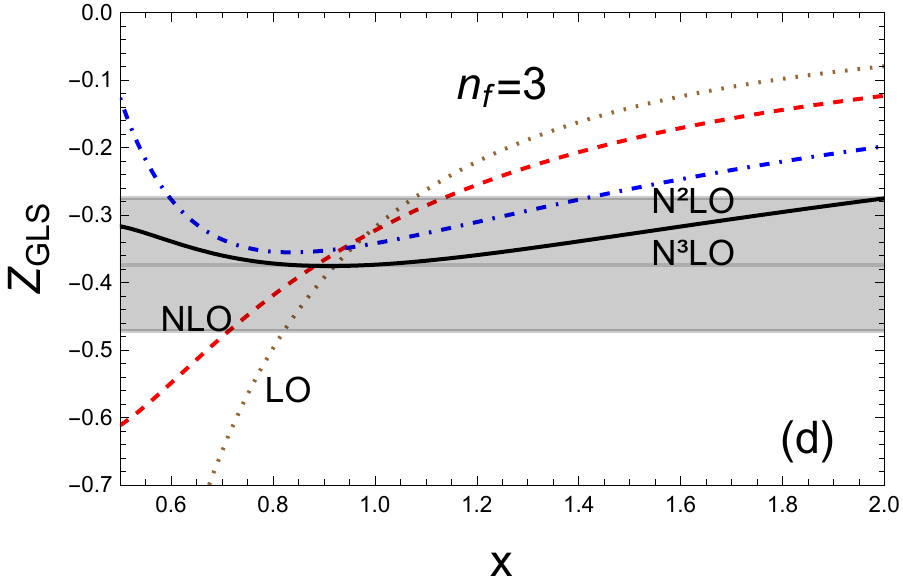}} &~
\hspace{25pt}
\subfigure[{$Z_{GLS}$} with $n_f=6$.]{\includegraphics[scale
=0.75]{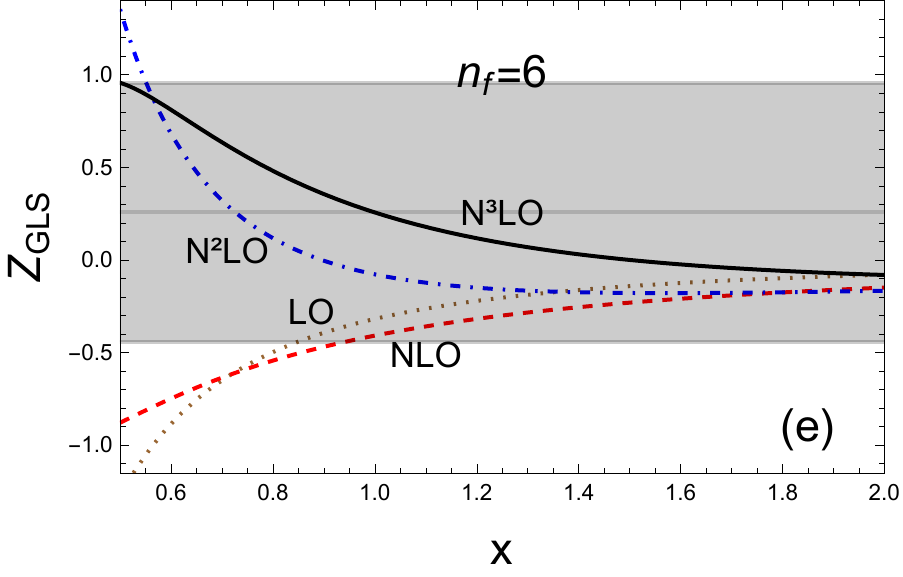}}
\\
\hspace{-20pt}
\subfigure[{$Z_{EJ}$} with $n_f=3$.]{\includegraphics[scale
=0.75]{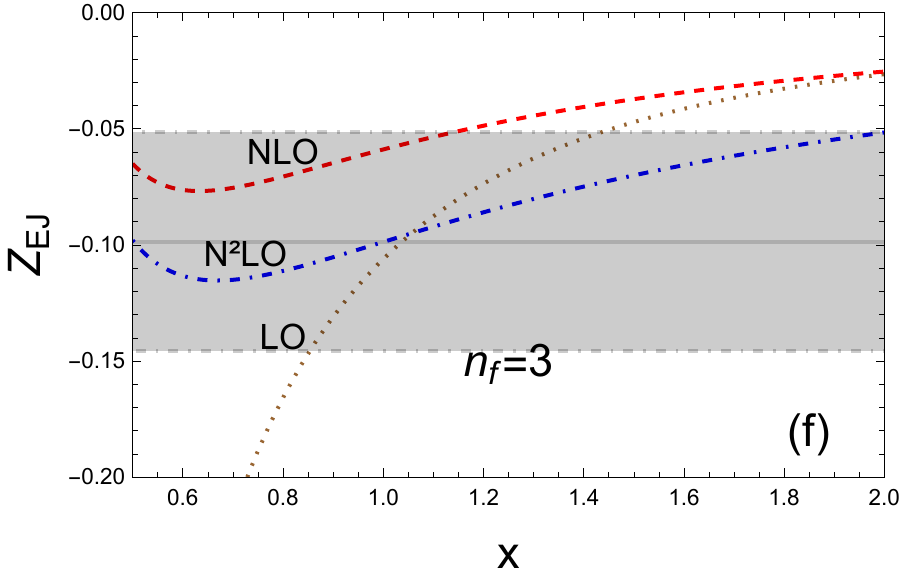}}&~
\hspace{25pt}
\subfigure[{$Z_{EJ}$} with $n_f=6$.]{\includegraphics[scale
=0.75]{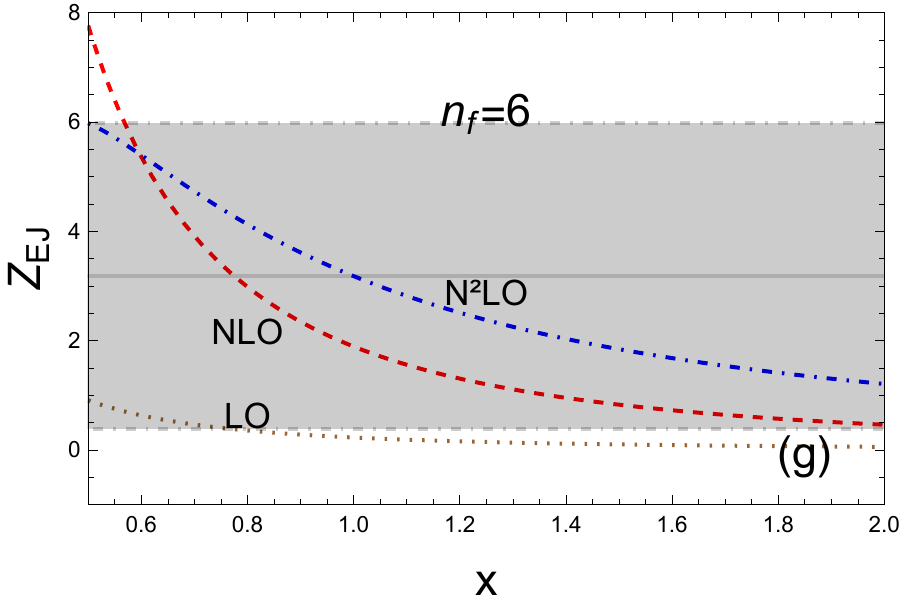}}\label{GLSnf6}
\end{tabular}
\caption {{\it Scale dependence of $Z_X$ for $n_f={0,3,6}$. 
The dotted (brown)
line is the LO result, the dashed (red) line is the NLO result, the dash-dotted (blue) line is the NNLO result, and the 
continuous (black) line the NNNLO result. The horizontal black line and the gray band represents our prediction and error quoted in Table~\ref{tableNX}.  
}}
\label{figNX}
\end{figure}

Our best values for $N_X$, and the associated error, can be found in 
Table~\ref{tableNX}. They have been computed with NNNLO accuracy for $Z_B$ and $Z_{GLS}$ (and with NNLO for $Z_{EJ}$, except for $n_f=0$, where all three sum rules are equal), after conformal mapping, 
at $x=1$. The scale dependence of the results, as well as the 
convergence, are shown in Fig.~\ref{figNX} for some selected values of $n_f$. 
In view of the figures, we believe that our error is conservative. To reinforce this conclusion, we also consider different possibilities for the expression of the asymptotic coefficient after the conformal mapping: either to do the conformal mapping to the original asymptotic expression, or to directly write what should be the new expression for the asymptotic expression (changing the location of the singularity). The differences are well inside the error. 

The error we quote in this paper is bigger than the error quoted in Ref. \cite{Campanario:2005np}. The reason is that we are more conservative here. If we had used the same method, we had obtained a larger error in that reference. Therefore, as we already said, we consider our error estimate to be conservative. On the other hand, the central values are quite close, except for large $n_f$, where we now observe a faster trend to get smaller values of the normalization as we increase $n_f$. 

In Fig. \ref{figNX}, we can see that the scale dependence becomes smother as we go to higher orders (for $n_f=0$ or 3). 
The convergence depends on the number of flavors. It is optimal for $n_f=3$, 
which actually happens to be the most interesting case 
from the physical point of view, and it deteriorates for large $n_f$. Also, for large $n_f$, the value of the normalization is consistent with zero.  This fits with the picture that 
the renormalon is less important when the number of flavors grows and one can 
reach to the point where the infrared renormalon disappears. For the Ellis-Jaffe 
perturbative series with $n_f\not=0$ (which is known to one order less), the same discussion applies but the flip of sign of the normalization happens at $n_f=4$, and for $n_f=6$, it starts to blow up. 

The values of $Z_B$ and $Z_{GLS}$ are consistent with each other within errors. 
This is consistent with the interpretation that the light-by-light term does 
not contribute to the renormalon, as it was done in Ref.~\cite{Contreras:2002kf}. 

We are then able to give some estimates for the coefficients of the perturbative 
series. We provide them in Tables~\ref{CB},~\ref{CGLS}, and~\ref{CEJS}. We should stress 
that our numbers incorporate the right asymptotic behavior, which is not the case for 
large-$\beta_0$ estimates. For the Bjorken and GLS sum rules, they were calculated 
in \cite{Broadhurst:1993ru}.

Previous estimates of the coefficients of the perturbative series can be found in the literature for the 
Bjorken (GLS) sum rules \cite{Kataev:1994gd,Kataev:1995vh,Ellis:1995jv,Contreras:2002kf}. Nevertheless, those estimates do not profit from the knowledge of the ${\cal O}(\al^4)$ coefficient, which was not known back then. 

\begin{table}[h!]
\hspace{-3cm}
\begin{center}
\begin{tabular}{|c|c|c|c|c|c|c|c|}
\hline
$n_f$&0 &1&2&3&4&5&6                                   \\\hline
-$C_{B}^{(0)}$& $0.506(186)$ &  $0.480(169) $ & $0.449(149)$ &$0.407(119)$ & $0.341(97)$ & $0.224(203)$ & $-0.04(0.44)$ \\\hline
-$C_{B}^{(1)}$ &$ 0.673(248)$ & $  0.589(208)  $ & $  0.502(166) $ & $0.407(119) $ & $ 0.295(84)  $ & $ 0.163(148) $ & $ -0.02(0.25)$ \\\hline
-$C_{B}^{(2)}$ & $1.48(55) $ & $ 1.207(426)$ & $ 0.947(314) $ & $0.698(204) $ & $ 0.453(129) $ & $  0.218(198)  $ & $ -0.02(0.28)  $ \\\hline
-$C_{B}^{(3)}$ & $4.57(1.68) $ & $ 3.471(1.22)  $ & $ 2.52(83) $ & $1.698(497) $ & $ 0.995(284) $ & $ 0.425(386) $ & $-0.04(0.48) $ \\\hline
-$C_{B}^{(4)}$ &$ 18.1(6.7) $ & $ 12.8(4.5) $ & $ 8.62(2.85) $ & $ 5.34(1.56) $ & $  2.85(0.81) $ & $  1.09(0.99) $ & $ -0.09(1.07) $ \\\hline
-$C_{B}^{(5)}$& $87.4(32.2) $ & $ 57.85(20.4) $ & $  36.1(12.0) $ & $ 20.6(6.0) $ & $  10.0(2.9) $ & $ 3.45(3.13) $ & $ -0.25(2.99)$ \\ \hline
-$C_{B}^{(6)}$& $499(184) $ & $ 309(109) $ & $  179.4(59.4) $ & $ 94.6(27.7) $ & $  42.0(12.0) $ & $ 13.0(11.8)  $ & $ -0.8(10.0)$ \\ \hline
\end{tabular}
\end{center}
\caption{\it Renormalon-based estimates of the perturbative coefficients 
$C_{B}^{(s)}(\nu)$ for $\nu=Q$ and for different number of flavors. We use the 
expression from Eq.~(\ref{CXRG}).}
\label{CB}
\end{table}
\begin{table}[h!]
\hspace{-3cm}
\begin{center}
\begin{tabular}{|c|c|c|c|c|c|c|}
\hline
$n_f$&1&2&3&4&5&6                                   \\\hline
-$C_{GLS}^{(0)}$&$ 0.473(165)$ & $ 0.432(138)$ & $0.374(98)$ & $ 0.279(158)$ & $ 0.113(323)$ & $ -0.256(700)$\\ \hline
-$C_{GLS}^{(1)}$& $ 0.580(203)$ & $ 0.482(154)$ & $ 0.373(98)$ & $ 0.243(137)$ & $ 0.083(235)$ & $ -0.146(399)$\\
\hline
-$C_{GLS}^{(2)}$& $ 1.19(42)$ & $0.910(291) $ & $ 0.640(168)$ & $ 0.372(211)$ & $ 0.111(315)$ & $ -0.164(449)$\\
\hline
-$C_{GLS}^{(3)}$& $ 3.42(1.19)$ & $ 2.42(77)$ & $ 1.56(41)$ & $ 0.818(463)$ & $ 0.215(614)$ & $ -0.28(76)$\\
\hline
-$C_{GLS}^{(4)}$& $ 12.6(4.4)$ & $8.28(2.64) $ & $ 4.90(1.29)$ & $ 2.34(1.32)$ & $ 0.55(1.57)$ & $ -0.62(1.70)$\\
\hline
-$C_{GLS}^{(5)}$& $ 57.0(19.9)$ & $ 34.7(11.1)$ & $18.9(5.0)$ & $ 8.24(4.66)$ & $ 1.75(4.98)$ & $ -1.7(4.7)$ \\ \hline
-$C_{GLS}^{(6)}$& $ 304(106)$ & $ 172(55)$ & $ 86.7(22.8)$ & $34.5(19.5)$ & $ 6.6(18.8)$ & $-5.8(15.9)$ \\ \hline
\end{tabular}
\end{center}
\caption{\it Renormalon-based estimates of the perturbative coefficients 
$C_{GLS}^{(s)}$ for $\nu=Q$ and for different number of flavors. We do not display 
the column with $n_f=0$ since the numbers are equal to the Bjorken case.
We use the 
expression from Eq.~(\ref{CXRG}).}
\label{CGLS}
\end{table}

\begin{table}[h!]
\hspace{-3cm}
\begin{center}
\begin{tabular}{|c|c|c|c|c|c|c|}
\hline
$n_f$&1&2&3              &4&5&6                    \\\hline
-$C_{EJ}^{(0)}$&$ 0.346(171)$ & $ 0.249(125)$ & $ 0.099(47)$
& $-0.176(106)$&$-0.816(485)$&$-3.18(2.80)$\\ \hline
-$C_{EJ}^{(1)}$&$ 0.406(201)$ & $ 0.252(126)$ & $ 0.083(40)$&-0.116(69)&-0.378(225)&-0.799(703)\\ \hline
-$C_{EJ}^{(2)}$& $ 0.810(402)$ & $ 0.449(225)$ & $ 0.129(61)$&-0.153(91)&-0.406(241)&-0.646(57)\\ \hline
-$C_{EJ}^{(3)}$& $ 2.28(1.13)$ & $ 1.14(57)$ & $ 0.293(140)$&-0.303(181)&-0.683(406)&-0.882(776)\\ \hline
-$C_{EJ}^{(4)}$& $ 8.32(4.13)$ & $ 3.80(1.90)$ & $ 0.877(418)$&-0.803(480)&-1.57(93)&-1.70(1.49)\\ \hline
-$C_{EJ}^{(5)}$& $ 37.1(18.4)$ & $ 15.5(7.8)$ & $ 3.25(1.55)$&-2.66(1.59)&-4.55(2.70)&-4.20(3.70)\\ \hline
-$C_{EJ}^{(6)}$& $ 196(97)$ & $ 75.4(37.8)$ & $ 14.4(6.9)$&-10.6(6.3)&16.0(9.5)&-12.8(11.2)\\ \hline
\end{tabular}
\end{center}
\caption{\it Renormalon-based estimates of the perturbative coefficients 
$C_{EJ}^{(s)}$ for $\nu=Q$ and for different number of flavors. We do not display 
the column with $n_f=0$ since the numbers are equal to the Bjorken case.
We use the 
expression from Eq.~(\ref{CXRG}). }
\label{CEJS}
\end{table}

\subsection{Hyperasymptotic expansion of $C_B^{\rm PV}$}
\label{PVsec}
\begin{figure}[ht]
     \begin{center}      
            \includegraphics[width=0.5\textwidth]{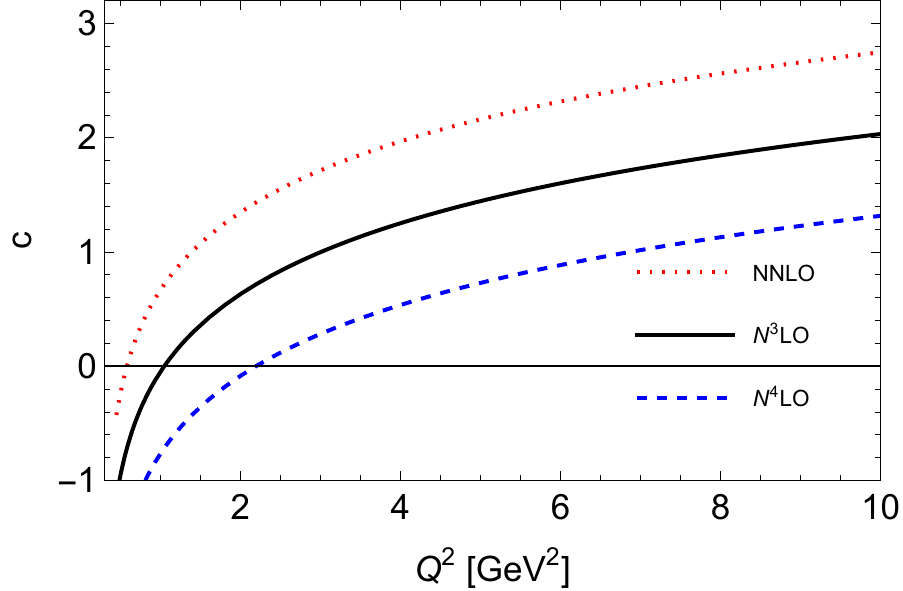} 
    \caption{ Value of $c$ from Eq. (\ref{NP}) as a function of $\mu^2=Q^2$. Red-dotted line is obtained fixing $N_P=2$, the black continuous line is obtained fixing $N_P=3$, and the dashed blue line is obtained fixing $N_P=4$.}
   \label{figC}
   \end{center}
\end{figure}

As we have already mentioned, the divergent behavior of the perturbative series is regulated using the PV prescription for the Borel sum [see Eq. (\ref{CBPV})]. The exact expression of the Borel transform is unknown. Therefore, one has to use approximations. In this context, it is of paramount importance to have a parametric control on the error with exponential accuracy. Consequently, we apply the hyperasymptotic approach developed in \cite{Ayala:2019uaw,Ayala:2019lak,Ayala:2019hkn}. The Borel sum can then be split into a partial sum truncated at the minimal term, plus the leading terminant, and plus a leftover of the perturbative expansion (the perturbative expansion is not known with high enough accuracy to go beyond that): 
\be
\label{CBPVhyp}
C_B^{\PV}(\als(Q))=C_B^{(N_P)}+\left[\frac{\als(Q)}{\als(\nu)}\right]^{-b_B}\Omega+
\sum_{n=N_P+1}^{N_{\rm max}}\left(C_B^{(n)}-C_B^{(n,\rm asym)}\right)\als^{n+1}(\nu)
\ee
where 
\be
\label{NP}
N_P=2 \frac{2\pi}{\beta_0 \als(\nu)}\left(1-c\als(\nu)\right)
\,,
\ee
\be
C_B^{(N_P)}(\als)=1+\sum_{n=0}^{N_P}C_B^{(n)}\als^{n+1}(\nu)
\,,
\ee
\be
\label{eq:Omegaexact}
\Omega \equiv Z_B\frac{\nu^{2}}{Q^{2}} \frac{1}{\Gamma(1+2 b+b_{B})}\left(\frac{\beta_{0}}{4 \pi}\right)^{N_P+1} \alpha_{X}^{N_P+2}(\nu) \int_{0, \mathrm{PV}}^{\infty} d x \frac{x^{2 b+b_B+N_P+1} e^{-x}}{1-x \frac{\beta_{0} \alpha_{X}(\nu)}{4 \pi }}
\,.
\ee
In the weak coupling limit, this expression can be approximated by
\be
\label{eq:Omega}
\Omega
\simeq K_{B,\MS}^{\PV}(c)\left[\als(\nu)\right]^{\frac{1}{2}-b_B}\frac{\Lambda_{\MS}^2}{Q^2}
\simeq K_{B,\MS}^{\PV}(c) \left[\als(\nu)\right]^{\frac{1}{2}-2b-b_B}
\left(\frac{\beta_0}{4\pi}\right)^{-2b}e^{-\frac{4\pi}{\beta_0\als(\nu)}}
\frac{\nu^2}{Q^2}
\ee
where
\be
\label{KBPV}
K_{B,\MS}^{\PV}(c)=-\frac{Z_B}{\Gamma(1+2b+b_B)}
\left(\frac{4\pi}{\beta_0}\right)^{b_B+1}\sqrt{\frac{\beta_0}{2}}\left[-\eta_c+\frac{1}{3}\right]
\,,
\ee
\be
\eta_c=-2b+\frac{4\pi}{\beta_0}c-b_B-1
\,.
\ee
$C_B^{(n,\rm asym)}$ is Eq. (\ref{CXRG}). $N_{\rm max}=3$. 
By default, we will take $N_P=N_{\rm max}$. This will fix $c$.  For illustration, we show the relation between $N_P$ and $c$ in Fig. \ref{figC}. In principle, we would like $c$ to be small or of order 1. For $Q$ around 1 GeV, $N_P=3$ is okay. For larger $Q$ one should ideally take a larger value for $N_P$. As $N_P=3$ is the maximum value known nowadays, it is not possible to take a largest value. In our analysis, we will always take $N_P=3$. We have checked that if we take $N_P=2$, and the corresponding $c$, the variation is minimal. 

Note the prefactor multiplying $\Omega$ in Eq. (\ref{CBPVhyp}). This characteristic is novel compared to earlier analyses using the hyperasymptotic expansion, and resums the large $\ln Q$ logarithms associated with the anomalous dimension. 

\begin{figure}[ht]
     \begin{center}      
            \includegraphics[width=0.5\textwidth]{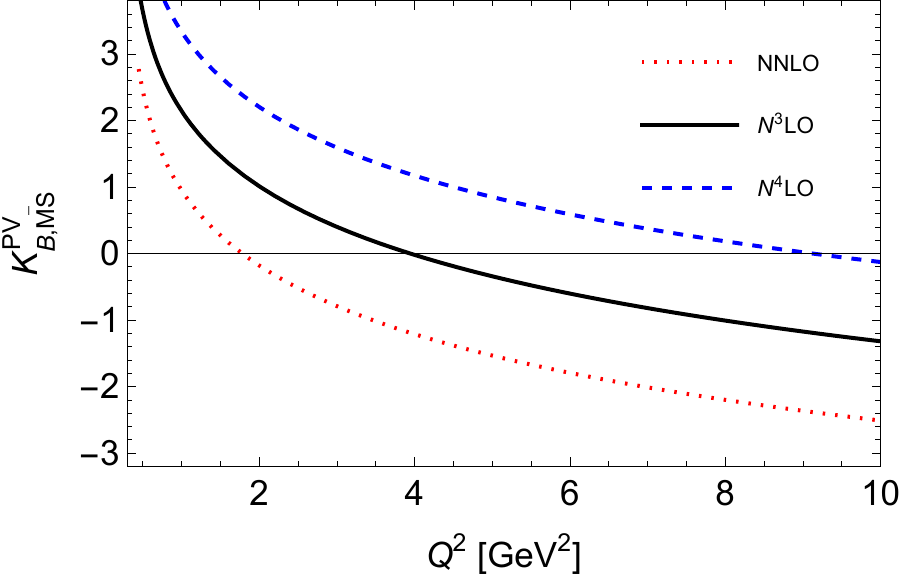} 
    \caption{ Value of $K_{B,\MS}^{\PV}$ from Eq. (\ref{KBPV}) as a function of $\mu^2=Q^2$. Red-dotted line is obtained fixing $N_P=2$, the black continuous line is obtained fixing $N_P=3$, and the dashed-blue line is obtained fixing  $N_P=4$.}
   \label{figK}
   \end{center}
\end{figure}

Note also that, in practice, $\Omega$ has a hidden dependence in $Q$ through $c$. In the weak coupling approximation, this hidden dependence is encoded in $K_B$. We show the dependence of $K$ on $Q^2$ in Fig. \ref{figK}. This makes that $\Omega$ effectively does not behave (leaving aside the anomalous dimension) as $\sqrt{\alpha(\mu)}\Lambda_{\MS}^2$. 

For numerics, we will use Eq. (\ref{eq:Omegaexact}). The difference with the weak coupling expressions in Eq. (\ref{eq:Omega}) is sizable for $Q^2 < 1$ GeV. We will discuss this issue further later.

\section{Comparison with experimental data and fit}

In this section, we will compare our theoretical predictions with the
experimental data. We
will use the experimental data for the Bjorken sum rule obtained in 
\cite{COMPASS:2016jwv,E143:1998hbs,SpinMuonSMC:1997mkb,E155:2000qdr,RSS:2006tbm,HERMES:2006jyl,Deur:2008ej,Deur:2004ti,Deur:2014vea,Deur:2021klh}.

We will combine the statistical and systematic errors of each experiment (in case they are given separately by the experiment) in quadrature. Nevertheless, for the more recent (and more precise) datasets from JLAB \cite{Deur:2008ej,Deur:2004ti,Deur:2014vea,Deur:2021klh}, some correlations are expected. To account for those, we follow the same procedure as in Refs. \cite{Deur:2014vea,Deur:2021klh} and combine part of the systematic error with the statistical one in quadrature, such that the reduced $\chi^2$ is fixed to be one. The remaining systematic error is taken to be completely correlated.

In the left-hand side of Eq.~(\ref{Bjorken}), target-mass effects have been included using the Nachtmann variable~\cite{Nachtmann:1973mr}, see also \cite{Piccione:1997zh,Blumlein:1998nv}. They read
\bea
    M_1^N(Q^2)
    &\equiv& \int_0^1 dx {\xi^{2} \over x^2} 
	\left\{ g_1^N(x, Q^2) \left[ {x \over \xi} - 
    {1 \over 9} {m_N^2 x^2 \over Q^2} {\xi \over x}
    \right] \right. 
	-\left. g_2^N(x, Q^2) ~ {m_N^2 x^2 \over Q^2} {4 \over 3} \right\} ~~~~
	\nonumber\\
    && 
	= \int_0^1 g_1^{N}(x,Q^2)dx + {\mu_4^N \over Q^2}+{\cal O}\left({1 \over Q^4}\right)
	= \Gamma_1^N(Q^2) + {\mu_4^N \over Q^2}+{\cal O}\left({1 \over Q^4}\right)
\,,
    \label{eq:i_nm1}
 \eea
where $N=p,n$ (remember that $M_1^B=M_1^p-M_1^n$, and accordingly $\Gamma_1^B=\Gamma_1^p-\Gamma_1^n$ and so on), $\xi = 2x / \left( 1 + \sqrt{1 + 4 m_N^2 x^2 / Q^2} \right)$ is the
Nachtmann scaling variable, $m_N$ is the nucleon mass. The quantity $M_1$
is the first Nachtmann moment of $g_1$ that absorbs all the
target mass corrections, $\sim (m_N^2/Q^2)^n$, and 
\begin{eqnarray}
\mu_4^{N}
&=& -\frac{m_N^2}{9}
    \left( a^{N}_2 + 4 d^{N}_2  \right),
\end {eqnarray}
where $a_2^{N}$ is the target mass correction given by the
$x^2$-weighted moment of the polarized leading-twist $g_1$ structure function,
and $d_2^{N}$ is a twist-three matrix element given by
\begin{eqnarray}
d^{N}_2
&=& \int_0^1 dx~x^2 \left( 2g^{N}_1 + 3g^{N}_2 \right).
\end{eqnarray}
For these quantities we use the values given in \cite{Deur:2014vea} (a recent lattice determination can be found in \cite{Burger:2021knd}):
\be
d_2^{p-n}=0.0080(36)\,, \qquad a_2^{p-n}=0.031(10)\,.
\ee

Note that the elastic contribution has to be included in the experimental
numbers in order the sum rule to be fully inclusive, i.e.,
\be
\Gamma_1^{p/n}(Q^2)={1 \over
2}F^{p/n}_1(Q^2)\left(F^{p/n}_1(Q^2)+F^{p/n}_2(Q^2)\right)+\Gamma_{\rm inel.}^{p/n}(Q^2)
\,.
\ee
The empirical parametrization of the elastic form factors is taken from the recent paper \cite{Sufian:2016hwn}. For large momentum this contribution is completely negligible. The difference starts to be sizable for $Q^2 < 2$ GeV$^2$. For $Q^2 <1$ GeV$^2$, it becomes very important. 

In comparing theory and experiment, we will let $\hat f_{3}^{\rm PV}$ to be a free parameter, which will then be fitted to the experimental data. The value of $\alpha_s(M_z)=0.0179(9)$ is taken from the PDG \cite{Tanabashi:2018oca}. After running down to scales of the order of 1 GeV using \cite{Herren:2017osy}, it yields 
\be
\label{eq:LMS}
\Lambda_{\MS}=335^{+14}_{-13}\; {\rm MeV}.
\ee
We assess the error associated with the parametrization of the elastic form factors by performing the same fits using the older parametrization \cite{Mergell:1995bf}. We then take the difference as the associated error. It will be small in comparison with the other experimental errors. 

In our final fits, we will neglect the error associated to the coefficients $g_A$, $a_2$ and $d_2$, as their impact is negligible in comparison with other uncertainties. 

\begin{figure}[ht]
     \begin{center}      
            \includegraphics[width=0.95\textwidth]{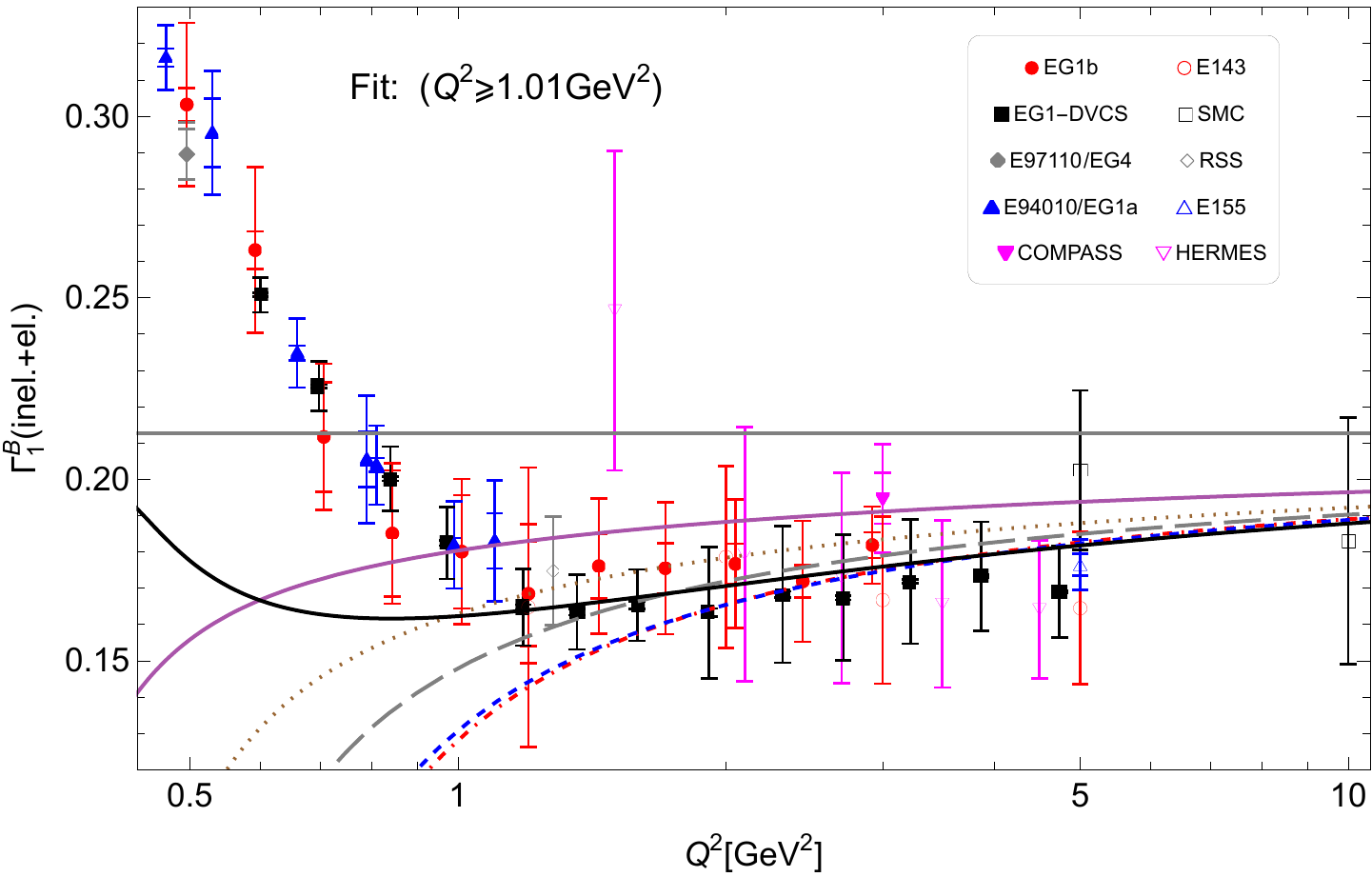} 
    \caption{{\it Leading-twist contribution to the sum rules at different orders in perturbation theory in the $\MS$ scheme 
with $\nu=Q$ compared with the experimental data.
The continuous horizontal gray line is $g_A/6$. The continuous magenta line is the LO result. The dotted brown line is the NLO result. The dashed gray line is the NNLO result. The dash-dotted red line is the NNNLO result. The blue dashed line corresponds to the NNNLO result plus the twist-two contribution. The continuous black line is the NNNLO result plus the twist-two contribution plus the terminant. }}
   \label{figMS}
   \end{center}
\end{figure}

We now consider the theory part of the problem. 
We first illustrate the problem of convergence of the perturbative
series by
drawing the perturbative series in the $\MS$ scheme at different orders in
$\als$ in Fig.~\ref{figMS}.

\begin{figure}[ht]
     \begin{center}      
            \includegraphics[width=0.95\textwidth]{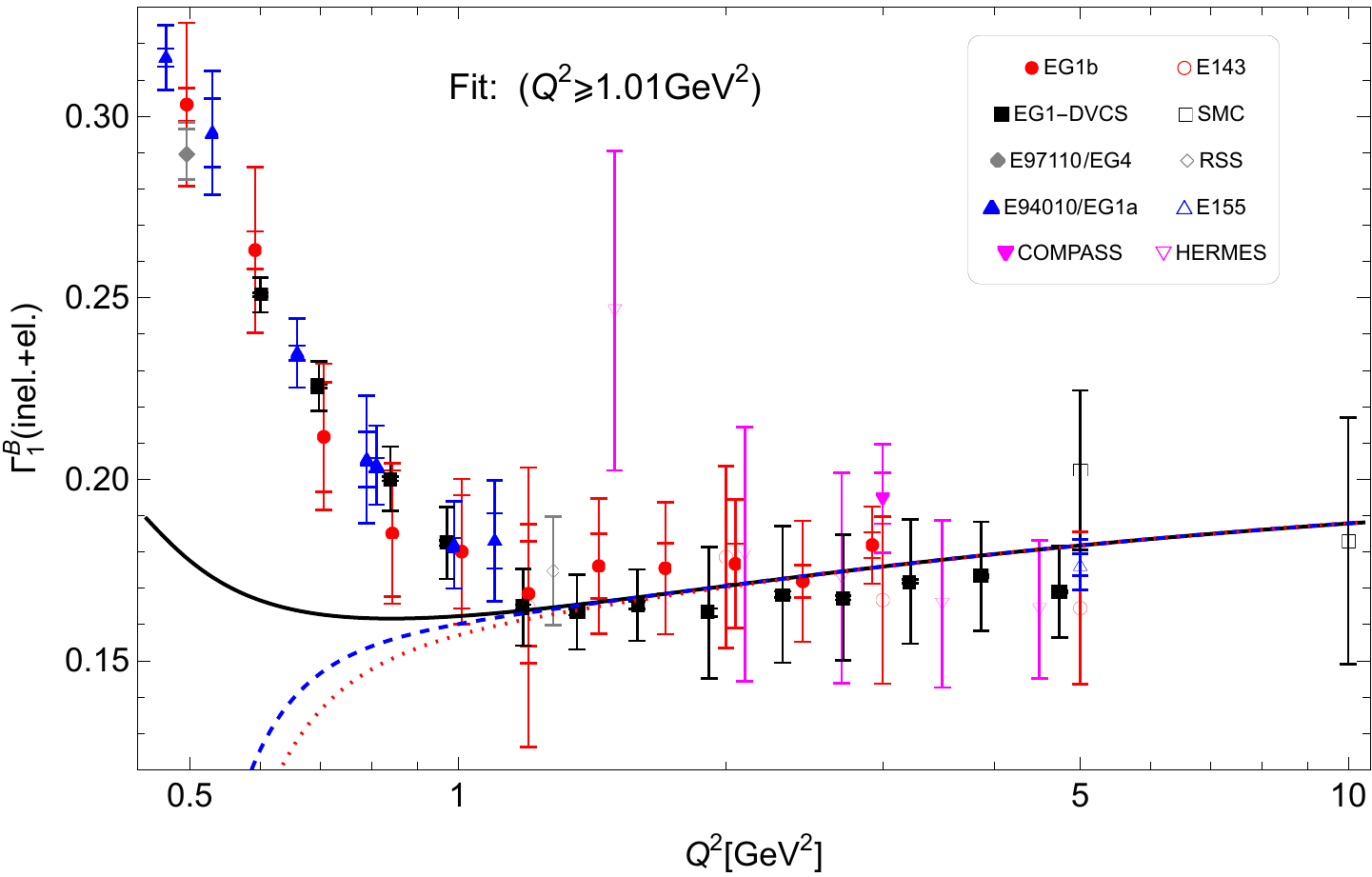} 
    \caption{\it The NNNLO result plus the twist-two contribution plus the terminant. The continuous-black line approximates the terminant to Eq. (\ref{eq:Omegaexact}), the dashed-blue line to the second equality of Eq. (\ref{eq:Omega}), and the dotted-red line to the first equality. }
   \label{FitOmegasamef3}
   \end{center}
\end{figure}
\begin{table}[ht]
\hspace{-3cm}
\begin{center}
\begin{tabular}{|c|c|c|}
\hline 
&Set I&Set II
\\
\hline
&$\hat f_{3,\MS}^{\PV}\times 10^3$
&$\hat f_{3,\MS}^{\PV}\times 10^3$
\\\hline
central
&32& 123
\\ \hline \hline
stat	
&$\pm$ 23 &$\pm$ 58
\\ \hline
sys	
&${}^{-174}_{+174}$ &  ${}^{-411}_{+411}$ 
\\ \hline
elastic	
&31&5
\\ \hline
$\Lambda_{\MS}^{(n_f=3)}$	
&${}^{-19}_{+19}$&${}^{-33}_{+33}$
\\ \hline \hline
$N_B$
&${}^{-52}_{+52}$ &  ${}^{-14}_{+14}$ 
\\ \hline
$\nu$	
&${}^{-4}_{-57}$ &  ${}^{+9}_{-20}$ 
\\ \hline
exp	
&-6&0
\\ \hline
\end{tabular}
\end{center}
\caption{\it 
Central values and error budget of the fit using the different datasets we use in this paper. In GeV$^2$, 
``stat'' stands for the statistical error of the fit, The others are self-explanatory. For further details see the main text. The fit uses Eq. (\ref{BjorkenPV}) with $N_P=3$ and Eq. (\ref{eq:Omegaexact}) for $\Omega$. The default value of $\nu$ is $Q$.}
\label{Tab:fitexpexp}
\end{table}
The perturbative series has a relative good
convergence.
However, this convergence deteriorates when we approach to low energies.
We also see how the
perturbative theoretical result diverges from the experimental numbers
at low energies. This is a reflection that, at these scales, we need a description of the experiment with exponential accuracy. In other words, power corrections have to be incorporated, and the perturbative sum needs to be evaluated with exponential accuracy. We do so by using Eq. (\ref{BjorkenPV}) with $C_B^{\PV}$ evaluated using the expression in Eq. (\ref{CBPVhyp}), which has exponential precision. We can see that the agreement with the experimental data improves. 
This is specially so around the 1 GeV region. We can also see a qualitative change in the figure for scales smaller than 1 GeV (see the black continuous line). Nevertheless, this is an artifact of using Eq. (\ref{eq:Omegaexact}) for the terminant. If we instead use the weak coupling expressions for it one has in Eq. (\ref{eq:Omega}), the behavior is different (see Fig. \ref{FitOmegasamef3}). This illustrates that we can not trust our predictions below 1 GeV. If we eliminate the terminant and only keep the dimension two condensate (blue dashed line in Fig. \ref{figMS}), the agreement deteriorates. The black line has been obtained from a fit letting $\hat f_{3}^{\PV}$ to be free using data fulfilling that $Q^2 \geq 1$ GeV$^2$. As just mentioned above, we have used Eq. (\ref{BjorkenPV}) with $C_B^{\PV}$ evaluated with the exponential precision using the expression in Eq. (\ref{CBPVhyp}) with $N_P=3$. For $\Omega$, we use Eq. (\ref{eq:Omegaexact}). The value obtained is listed in Table \ref{Tab:fitexpexp}. 

We now turn to the discussion of how robust this result is, and to determine the error. In Table \ref{Tab:fitexpexp}, we disclose in detail the error budget. 

The experimental errors have already been discussed above. We have separated them into the statistical, systematic, and the one associated with different parametrizations of elastic terms. By far, the dominant error is the systematic error associated with the JLAB data. We also include the error due to $\al_s$ (see Eq. (\ref{eq:LMS}) in Table \ref{Tab:fitexpexp}), which is also much smaller than the systematic error. We then combine them in quadrature to give our final experimental error. 

The theoretical error is generated by our incomplete knowledge of $C_B^{\PV}$ and the incomplete knowledge of the Wilson coefficient of the dimension-two term of the OPE. Theoretically, there is also the error associated with higher order terms of the OPE. These are of ${\cal O}(1/Q^4)$. This is parametrically smaller than the other errors associated with our incomplete knowledge of the terminant and of the Wilson coefficient of the higher twist. The error of  $C_B^{\PV}$ is parametrically of the order ${\cal O}(\als^{3/2-2b-b_B}e^{-\frac{4\pi}{\beta_0\als}})$. The error of the higher twist term is parametrically of the order ${\cal O}(\als^{1-2b-b_B}e^{-\frac{4\pi}{\beta_0\als}})$. We estimate this error using a variety of methods:

 \begin{enumerate}[label=(\arabic*)]
\item We allow for a variation of
 $N_B$ according to the error given in Table~\ref{tableNX}.
\item We consider the $\nu$ scale dependence of the result and allow for a variation of $\nu^2 \in (Q^2/2,2Q^2)$.
\item Approximating $\Omega$ by its weak coupling expansion at leading order. In other words, working with the second equality in Eq. (\ref{eq:Omega}).
 \end{enumerate}
We now discuss these methods to estimate the error in more detail.
 \begin{enumerate}[label=(\arabic*)]
\item  The error of $N_B$ is due to our incomplete knowledge of the perturbative series, and it can be, to some extent, be mixed with the ${\cal O}(\als^{3/2}e^{-\frac{4\pi}{\beta_0\als}})$ corrections of the terminant. 

\item The scale dependence is a reflection of the incomplete knowledge of the terminant and of the Wilson coefficient of  the dimension two term of the OPE. If we keep the running of the coupling constant consistently with the precision of the computation, there would not be scale dependence. We implement the scale dependence by considering the running of the coupling with the maximal known accuracy.  

\item Measures the size of the ${\cal O}(\als^{3/2}e^{-\frac{4\pi}{\beta_0\als}})$ correction to the terminant. An alternative way to measure these corrections is to approximate $\Omega$ using the first equality of Eq. (\ref{eq:Omega}). This indeed produces larger errors but still much smaller than the error associated with $N_B$, or $\nu$. Another alternative consists of changing $N_P$ from 3 to 2. This changes $c$. If the terminant were known with infinite precision, the dependence in $c$ would disappear, in the sense that the dependence on $c$ of the terminant would cancel with the dependence on $c$ of the finite sum. Therefore, the variation of $c$ assesses such dependence. The effect happens to be very small.
\end{enumerate}
Let us now come back to the issue of the ${\cal O}(1/Q^4)$ corrections. As we have already mentioned, they are parametrically subleading than the errors we have considered. Therefore, we will neglect them.\footnote{Their inclusion may improve the fit \cite{Deur:2014vea,Ayala:2018ulm}, but it is in potential conflict with neglecting ${\cal O}(\alpha_s)$ suppressed effects of the leading power suppressed corrections, and with neglecting subleading renormalons of the perturbative series. Therefore, we refrain from doing so in this paper.} Still, fits with different energy ranges may give indirect hints of their possible importance. Related to this discussion, a potentially worrisome issue is the size of the elastic terms. For large momentum this contribution is completely negligible. The difference starts to be sizable for $Q^2 < 2$ GeV$^2$. For $Q^2 <1$ GeV$^2$ becomes very important. In this respect, it is important to mention that the short distance behavior of the elastic form factors is expected to produce corrections to the Bjorken sum rule with powers bigger than $1/Q^2$ \cite{Lepage:1980fj} (for the present discussion, see also \cite{Mergell:1995bf,Sufian:2016hwn}). Therefore, one may argue that they are a measure of power corrections of order $1/Q^4$ or higher. Nevertheless, such prediction is not in such a robust status as fully inclusive sum rules, as they somehow rely on local quark-hadron duality (one expects them to fail in the large $N_c$ limit). Therefore, we will not consider this as a source of error, although it would deserve further studies.

For the final estimate of the theory error, we consider items 1-3 as different estimators of the incomplete knowledge of the OPE. Even though, there is some overlapping between them (they measure similar things), we will consider them as independent, and combine them in quadrature to give our final theory error of $\hat f_{3}^{\rm PV}$.

The sensitivity to the theoretical and experimental error may vary depending on the range of energies used for the experimental data. Therefore, we use different datasets for the fits to see the dependence on the dataset. The different datasets we consider are the following: set I: $Q^2 \geq 1$ GeV$^2$ and set II: $Q^2 \geq 2$ GeV$^2$. The set I has 31 points, and the set II has 19 points (and are less precise). We refrain from considering smaller values of $Q$, since the dependence on the specific parametrization of $\Omega$ becomes important. As a general trend, when we increase the number of points, the statistical and systematic errors of $\hat f_{3}^{\rm PV}$ decrease. On the other hand, the error associated with the parametrization of the elastic form factors moderately increases but to a lesser extent. On the theory side, we observe the following dependence on the dataset: As expected, the fit is more dependent on $N_B$, as we go to smaller energies (set I). We find a large dependence on $\nu$ with the fit to the set I if we lower $\nu$ since we take $\nu^2=Q^2/2$. On the other hand, we find very little dependence if we take higher values of $\nu$. For set II the error is more symmetric and, overall, smaller. Finally, the error associated with $\Omega$ is very tiny, and basically zero when using set II for the fit.

For our final numbers we use the results obtained with the dataset I. This dataset yields smaller errors. For convenience, we display the results obtained for the different datasets:

\bea
{\rm Set\; I}:\quad \hat f_{3}^{\PV}\times 10^3
&=&32^{+180}_{-180}(\rm exp)^{+52}_{-77}(\rm th) 
=32^{+187}_{-196} \;{\rm GeV}^2 ,
\\
{\rm Set\; II}:\quad \hat f_{3}^{\PV}\times 10^3&=&123^{+417}_{-417}(\rm exp)^{+17}_{-24} (\rm th) =123^{+417}_{-418}\;{\rm GeV}^2 .
\eea
For the final error, we combine the experimental and theoretical error in quadrature. The experimental error is the dominant one. As we have already mentioned, we take the numbers obtained from the fit to the dataset I as our final numbers. Our final number for $\hat f_{3}^{\PV}$ reads:
\be
\hat f_{3}^{\PV}\times 10^3=32^{+187}_{-196}\;{\rm GeV}^{2}
\,.
\ee
We also give a number for $\bar f_{3,\MS}^{\PV}$ [defined in Eq. (\ref{eq:hatf3PV})]. Its error is overwhelmingly dominated by the error of $\hat f_{3}^{\PV}$, being the error associated to $\Lambda_{\MS}^{n_f=3}$ way subleading. 
We obtain
\be
\bar f_{3,\MS}^{\PV}=0.29^{+1.67}_{-1.75}\;
\,.
\ee

\begin{figure}[ht]
     \begin{center}      
            \includegraphics[width=0.9\textwidth]{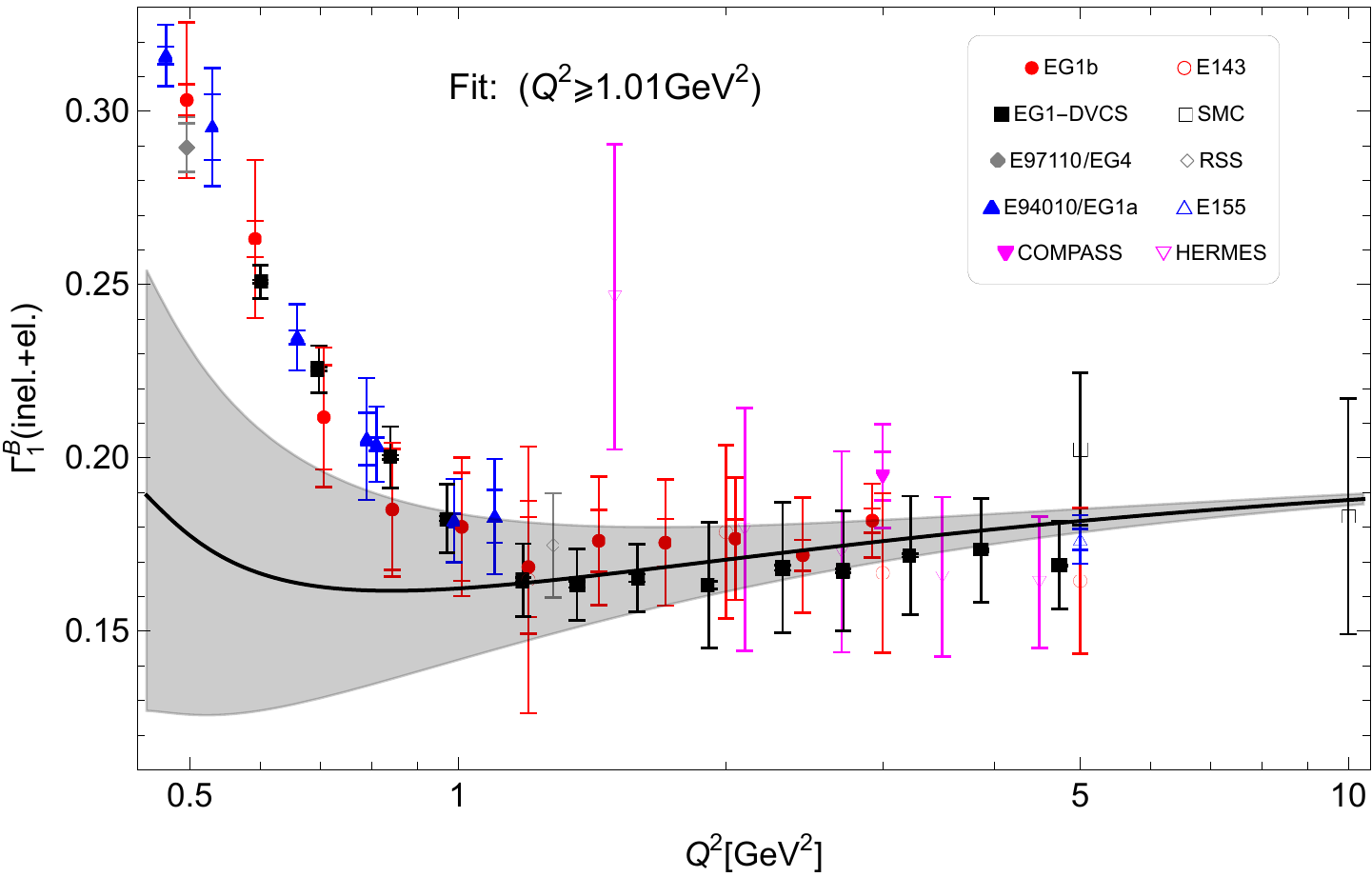} 
    \caption{{\it Plot of $\Gamma_1^{B}$ black line: fitted line with the value $\hat f_{3,\MS}^{\PV}=0.032^{+0.187}_{-0.196}$. The gray band is the $\hat f_{3,\MS}^{\PV}$ associated error.}}
   \label{figErrorf3}
   \end{center}
\end{figure}

\begin{figure}[ht]
     \begin{center}      
            \includegraphics[width=0.9\textwidth]{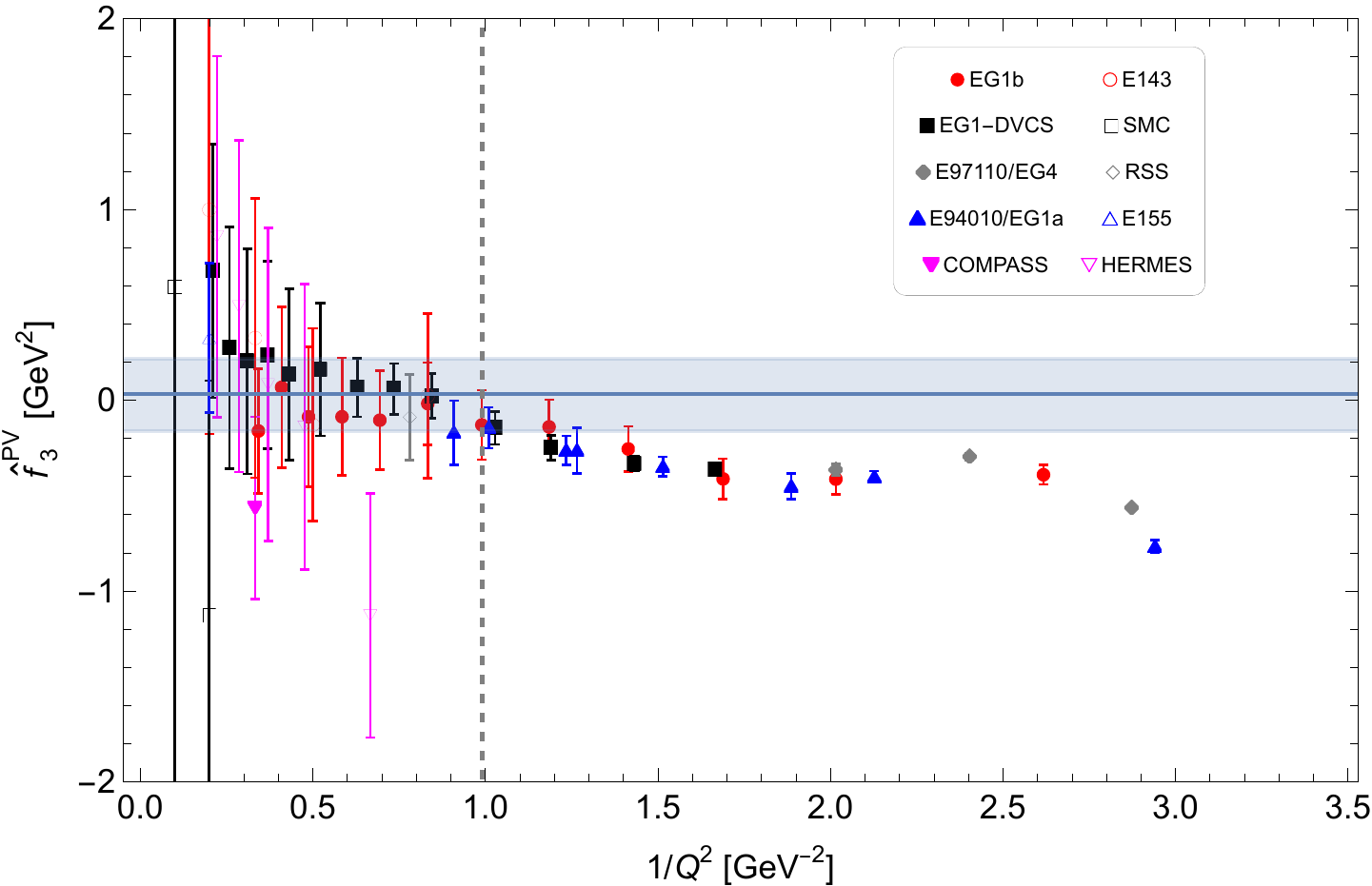} 
    \caption{ {\it Plot of $\hat f_{3}^{\PV}=0.032^{+0.187}_{-0.196}$ GeV$^2$. The gray band is the $\hat f_{3,\MS}^{\PV}$ associated error. }}
   \label{fighatf3}
   \end{center}
\end{figure}

To illustrate the quality of the fit and the impact of the error, we plot our final results in different ways. In Fig.  \ref{figErrorf3}, we plot our final predicted curve for $\Gamma_1^B$, including the error associated with $\hat f_{3}^{\PV}$. In Fig. \ref{fighatf3}, we directly show our prediction of $\hat f_{3}^{\PV}$ (and its associated error) and how it compares with the experimental data. We observe that 
the fit (the experimental data) approximately follows a constant value for $\hat f_3^{\PV}$, as expected from the use of a summation scheme of the perturbative series that is scheme and scale independent.

\medskip

The results obtained in this paper can be compared with the more traditional renormalon subtracted (RS) scheme \cite{Pineda:2001zq}, as used in Ref. \cite{Campanario:2005np}. We obtain $f_{\RS}(1;{\rm GeV})=0.152$ GeV$^2$, $f_{\RS}(0.8;{\rm GeV})=0.105$ GeV$^2$ using the dataset I and $f_{\RS}(1;{\rm GeV})=0.260$ GeV$^2$, $f_{\RS}(0.8;{\rm GeV})=0.239$ GeV$^2$ using the dataset II. 

\begin{figure}[ht]
     \begin{center}      
            \includegraphics[width=0.9\textwidth]{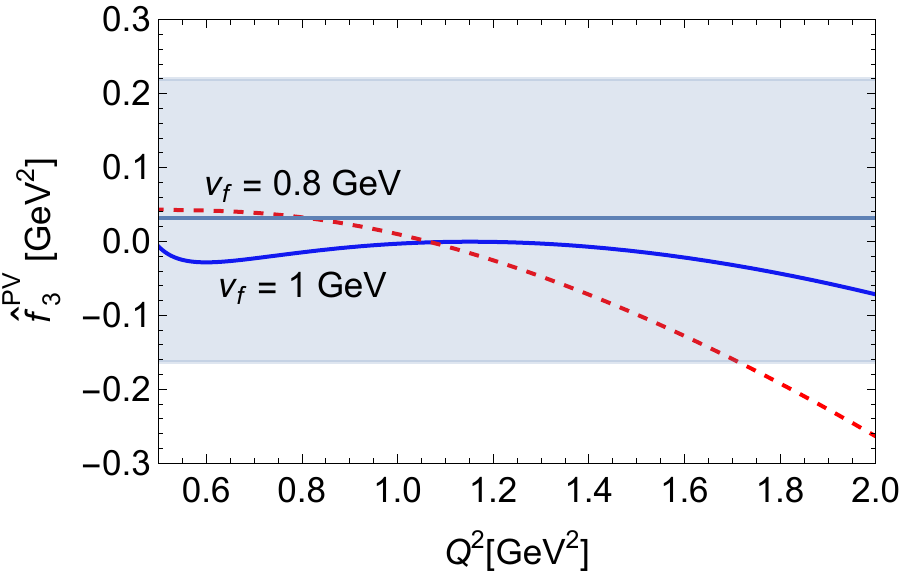} 
    \caption{ { Plot of $\hat f_{3}^{\PV}$ obtained from the RS determination using dataset I for different values of $Q^2$ and for the value of $\nu_f=1$ and 0.8 GeV. The horizontal blue line and gray band is the value of $\hat f_{3,\MS}^{\PV}$ and its associated error of our primary fit. }}
   \label{fighatf3RS}
   \end{center}
\end{figure}

To translate these determinations of $f_3$ from the RS scheme to the PV scheme, one has to use the following expression ($N_P=N_{\rm max}$ and $Q=\nu$):
\be
\hat f_{3}^{\PV}
=
f_{3,\RS}(\nu_f)\left[\als(\nu_f)\right]^{-\frac{\gamma_{\rm NS}^0}{2\beta_0}}+
\frac{27}{24}Q^2\left[\als(Q)\right]^{-\frac{\gamma_{\rm NS}^0}{2\beta_0}}
g_A
\left(
\sum_{s=1}^{N_P}
C_B^{s,asym}(\nu_f;Q)\als^{s+1}(Q)+\Omega(N_P;\nu=Q)
\right)
\,.
\ee 
Note that the left-hand side is independent of $Q$ and $\nu_f$. If we take  $\nu_f=Q=1$ GeV in the right-hand side, then
the fits with the RS scheme give very similar numbers to the ones we have obtained in this paper. Using set I, one obtains $\hat f_{3}^{\PV}=-0.003$ GeV$^2$, and with set II, one obtains $\hat f_{3}^{\PV}=0.141$ GeV$^2$. This shows the stability of the fit to using a different scheme for handling the renormalon. Within errors, the result is stable to variation of $\nu_f$ and $Q$, as shown in Fig. \ref{fighatf3RS}, for dataset I.

We can also compare with the result obtained in \cite{Campanario:2005np} using the RS scheme. We only aim to make a qualitative comparison. In that paper, it was obtained $f_{3,\RS}(\nu_f)=-0.124^{+0.137}_{-0.142}$ by a global fit to the Bjorken, Ellis-Jaffe, and the GLS sum rules to existing data back then with $Q^2 \geq 0.66$ GeV$^2$. A pure fit to the very same data to only the Bjorken sum rule (and treating slightly different the strong coupling) yields -0.102 GeV$^2$. If we use the new experimental data, the new value of $N_B$, and the new known coefficient of the perturbative expansion, then the result does not change much, one obtains -0.091 GeV$^2$. Note that all the experimental systematic error has been combined with the statistical one in quadrature, since the reduced $\chi^2$ is 1.6, bigger than 1. Actually, this is the reason the result is similar to the one obtained in\cite{Campanario:2005np}, as it makes that the new data weights similarly to the old one. If one keeps part of the systematic error as independent, then the new data weighs more, and shifts the central value (but then the rest of the error goes into the systematics, not affecting the comparison within errors). This effect is seen in the fits with data constrained to be $Q^2 \geq 1$ GeV$^2$ or $Q^2 \geq 2$ GeV$^2$. Taking into account the error of the determinations and of the conversion, the difference is of the order of one sigma, which we consider reasonable. Moreover, in this paper, we prefer to play a more conservative attitude, in view of the different behavior of the terminant (depending on the approximation used for it) for $Q^2 \leq 1$, and restrict to fits with $Q^2 \geq 1$ GeV$^2$.

\section{Conclusions}

In this paper, an updated determination of the normalization of the leading infrared renormalon of the Bjorken, GLS and Ellis-Jaffe sum rule has been obtained. Compared with the evaluation of Ref. \cite{Campanario:2005np}, this evaluation profits from new terms of the perturbative expansion that are known now, as well as from a more optimal way to pin down the renormalon, as highlighted in Ref. \cite{Bali:2013pla}. The results are summarized in Table \ref{tableNX}. For the most relevant cases of the Bjorken sum rule for $n_f=0$ and $n_f=3$, they read
\be
N_B(n_f=0)=-0.506 \pm 0.186\;, \qquad N_B(n_f=3)=-0.407\pm 0.119
\,.
\ee
Using these results, estimates of higher order terms of the perturbative series are also given (see Tables \ref{CB}, \ref{CGLS}, and \ref{CEJS}). 

For the first time ever, the hyperasymptotic approximation has been applied and validated directly in experimental data (up to possible concerns on the extrapolations to $x \rightarrow 0$ of the dispersive integrals of the sum rules, where there is no direct experimental data). We computed the 
Bjorken sum rule with hyperasymptotic precision by including the leading terminant, associated with the first infrared renormalon. One advantage of the present method is that it minimizes the dependence on the normalization of the renormalon. Another major advantage is that it allows us to obtain the ``real'' size of the nonperturbative corrections, in the sense that the nonperturbative correction scales as powers of $\Lambda_{QCD}$. 

After fitting the experimental data to the OPE theoretical expression letting $\bar f_{3,\MS}^{\PV}$ as a free parameter, we obtain a reasonable agreement with experiment for $Q^2 \geq 1$ GeV$^2$ with 
\be
\bar f_{3,\MS}^{\PV}=0.29^{+1.67}_{-1.75}\;,
\qquad
\hat f_{3}^{\PV}=0.032^{+0.187}_{-0.196}\;{\rm GeV}^{2}
\,.
\ee
We emphasize that this result is independent of the renormalization scale and scheme used for the strong coupling. The error of $\bar f_{3,\MS}^{\PV}$ is overwhelmingly dominated by the error of $\hat f_{3}^{\PV}$, and the latter is overwhelmingly dominated by the experimental error. There is some correlation in the experimental error of the most precise data from the JLAB experiments \cite{Deur:2008ej,Deur:2004ti,Deur:2014vea,Deur:2021klh}. We follow the methodology of \cite{Deur:2014vea,Deur:2021klh}. This correlation yields the most important source of error of the present analysis from the experimental side. Any improvement in this respect will immediately lead to a reduction of the errors of the numbers obtained in this paper.

As we can see in Figs. \ref{figMS} and \ref{fighatf3}, the fit complies  with expectations. The difference between the leading twist contribution and the experimental result is reasonably fitted by a quadratic correction with a $Q$-independent value of $\hat f_{3}^{\PV}$. It approximately follows a straight line (see Fig. \ref{fighatf3}), as expected.  Its value is small and consistent with zero within one sigma significance. 

The introduction of the terminant makes the result more stable. The inclusion of the 
terminant introduces a qualitative change in the perturbative behavior around
the 1 GeV region, making it much closer to the experimental figure. One may think that the terminant is basically equivalent to the nonperturbative condensate (with the same anomalous dimension), albeit $\sqrt{\alpha(\mu)}$ suppressed. Note, however, that, effectively, $\Omega$ has an implicit $Q$ dependence, since it depends on $c$, which is fixed such that $N_P=3$. 

The theoretical error of $\hat f_{3}^{\PV}$ is dominated by the scale variation and $N_B$ and $\nu$, and it is bigger when fitting to the dataset I than to the dataset II. In any case, it is still much smaller than the experimental error. All these theoretical errors would benefit from higher order computations. To know the anomalous dimension of the condensate to a higher order would also be of help. 

We let to future research the application of the hyperasymptotic approximation to other sum rules. 

\medskip
   
\noindent
{\bf Acknowledgments.}\\
 We would like to thank A. Deur for correspondence. This work was supported in part by FONDECYT (Chile) Grant No. 1200189, and by the Spanish Ministry of Science and Innovation 
 (Grant No. PID2020-112965GB-I00/AEI/10.13039/501100011033). This project has received funding from the European Union's Horizon 2020 research and innovation programme under Grant Agreement No. 824093. IFAE is partially funded by the CERCA program of the Generalitat de Catalunya. 

\appendix



\begin{thebibliography}{99}

\bibitem{Bjorken}
J. D. Bjorken,
Phys. Rev. \textbf{148}, 1467 (1966).

\bibitem{Shifman:1978bx}
M.~A.~Shifman, A.~I.~Vainshtein and V.~I.~Zakharov,
Nucl. Phys. B \textbf{147}, 385-447 (1979)

\bibitem{Campanario:2005np}
F.~Campanario and A.~Pineda,
Phys. Rev. D \textbf{72}, 056008 (2005)
[arXiv:hep-ph/0508217 [hep-ph]].

\bibitem{bj1loop}
  J. Kodaira, S. Matsuda, T. Muta, K. Sasaki, T. Uematsu,
  Phys. Rev. D {\bf 20} (1979) 627;\\
  J. Kodaira, S. Matsuda, K. Sasaki, T. Uematsu, Nucl. Phys. B {\bf 159}
  (1979) 99.
\bibitem{bj2loop}
   S.G. Gorishny, S.A. Larin, Phys. Lett. B {\bf 172} (1986) 109.
\bibitem{bj3loop}
   S.A. Larin, J.A.M. Vermaseren, Phys. Lett. B {\bf 259} (1991) 345.
   
\bibitem{Baikov:2010je}
P.~A.~Baikov, K.~G.~Chetyrkin and J.~H.~Kuhn,
Phys. Rev. Lett. \textbf{104}, 132004 (2010)
[arXiv:1001.3606 [hep-ph]].

\bibitem{Larin:2013yba}
S.~A.~Larin,
Phys. Lett. B \textbf{723}, 348-350 (2013)
[arXiv:1303.4021 [hep-ph]].

\bibitem{Baikov:2010iw}
P.~A.~Baikov, K.~G.~Chetyrkin and J.~H.~Kuhn,
Nucl. Phys. B Proc. Suppl. \textbf{205-206}, 237-241 (2010)
[arXiv:1007.0478 [hep-ph]].

\bibitem{Blumlein:1998sh}
J.~Blumlein and W.~L.~van Neerven,
Phys.\ Lett.\ B {\bf 450}, 417 (1999)
[arXiv:hep-ph/9811351].

\bibitem{Blumlein:2016xcy}
J.~Bl\"umlein, G.~Falcioni and A.~De Freitas,
Nucl. Phys. B \textbf{910} (2016), 568-617
[arXiv:1605.05541 [hep-ph]].
\bibitem{Ayala:2014yxa}
C.~Ayala, G.~Cveti\v{c} and A.~Pineda,
JHEP \textbf{09}, 045 (2014)
[arXiv:1407.2128 [hep-ph]].

 \bibitem{highertwists}
   V.M. Braun, A.V. Kolesnichenko, Nucl. Phys. B {\bf 283} (1987) 723.
   
\bibitem{powercorrections}
   I.I. Balitsky, V.M. Braun, A.V. Kolesnichenko, Phys. Lett. B {\bf 242}
  (1990) 245; E-ibid  B {\bf 318} (1993) 648.

\bibitem{Kawamura:1996gg}
H.~Kawamura, T.~Uematsu, J.~Kodaira and Y.~Yasui,
Mod.\ Phys.\ Lett.\ A {\bf 12}, 135 (1997)
[arXiv:hep-ph/9603338].

\bibitem{ParticleDataGroup:2020ssz}
P.~A.~Zyla \textit{et al.} [Particle Data Group],
PTEP \textbf{2020}, no.8, 083C01 (2020).

\bibitem{Parisi:1978bj}
  G.~Parisi,
  Phys.\ Lett.\ B {\bf 76}, 65 (1978).

\bibitem{Mueller:1993pa}
  A.~H.~Mueller,
  Phys.\ Lett.\ B {\bf 308}, 355 (1993).

\bibitem{tHooft}
G. 't Hooft, in {\it The Whys of Subnuclear Physics}, edited by A. Zichichi 
(Plenum, New York, 1978).

\bibitem{Ayala:2019uaw}
C.~Ayala, X.~Lobregat and A.~Pineda,
Phys. Rev. D \textbf{99}, no.7, 074019 (2019)
[arXiv:1902.07736 [hep-th]].

\bibitem{Takaura:2020byt}
H.~Takaura,
JHEP \textbf{10}, 039 (2020)
[arXiv:2002.00428 [hep-ph]].

\bibitem{Bali:2014sja}
G.~S.~Bali, C.~Bauer and A.~Pineda,
Phys. Rev. Lett. \textbf{113}, 092001 (2014)
[arXiv:1403.6477 [hep-ph]].

\bibitem{Ayala:2020pxq}
C.~Ayala, X.~Lobregat and A.~Pineda,
JHEP \textbf{12}, 093 (2020)
[arXiv:2009.01285 [hep-ph]].

\bibitem{Ayala:2019lak}
C.~Ayala, X.~Lobregat and A.~Pineda,
Nucl. Part. Phys. Proc. \textbf{309-311}, 77-86 (2020)
[arXiv:1910.04090 [hep-ph]].

\bibitem{Ayala:2019hkn}
C.~Ayala, X.~Lobregat and A.~Pineda,
Phys. Rev. D \textbf{101}, no.3, 034002 (2020)
[arXiv:1909.01370 [hep-ph]].

\bibitem{BerryandHowls}
M. V. Berry and C. J. Howls, Hyperasymptotics, Proc. Roy. Soc. London
A, 430 (1990), pp. 653-668.
  
\bibitem{Dingle} R.B. Dingle, {\it Asymptotic Expansions: Their
Derivation and Interpretation} (Academic Press, London, 1973).

\bibitem{Ayala:2020odx}
C.~Ayala, X.~Lobregat and A.~Pineda,
JHEP \textbf{09}, 016 (2020)
[arXiv:2005.12301 [hep-ph]].

\bibitem{Bali:2013pla}
G.~S.~Bali, C.~Bauer, A.~Pineda and C.~Torrero,
Phys. Rev. D \textbf{87}, 094517 (2013)
[arXiv:1303.3279 [hep-lat]].

\bibitem{GLS69} D.J. Gross, C.H. Llewellyn Smith, 
 Nucl. Phys. B {\bf 14}, 337 (1969).

\bibitem{EllJaf74}J. Ellis, R.L. Jaffe,  Phys. Rev {\bf D} 9, 1444 (1974), 
E-ibid {\bf 10}, 1669 (1974). 

\bibitem{Blumlein:2012bf}
J.~Bl\"umlein,
Prog. Part. Nucl. Phys. \textbf{69} (2013), 28-84
[arXiv:1208.6087 [hep-ph]].

\bibitem{Beneke:1997qd}
M.~Beneke, V.~M.~Braun and N.~Kivel,
Phys.\ Lett.\ B {\bf 404}, 315 (1997)
[arXiv:hep-ph/9703389].

\bibitem{Contreras:2002kf}
C.~Contreras, G.~Cvetic, K.~S.~Jeong and T.~Lee,
Phys.\ Rev.\ D {\bf 66}, 054006 (2002)
[arXiv:hep-ph/0203201].

\bibitem{Ellis:1995jv}
  J.~R.~Ellis, E.~Gardi, M.~Karliner and M.~A.~Samuel,
  Phys.\ Lett.\ B {\bf 366}, 268 (1996)
  [arXiv:hep-ph/9509312].

\bibitem{Broadhurst:1993ru}
D.~J.~Broadhurst and A.~L.~Kataev,
Phys. Lett. B \textbf{315}, 179-187 (1993)
[arXiv:hep-ph/9308274 [hep-ph]].

\bibitem{Kataev:1994gd}
A.~L.~Kataev,
Phys.\ Rev.\ D {\bf 50}, 5469 (1994)
[arXiv:hep-ph/9408248].

\bibitem{Kataev:1995vh}
A.~L.~Kataev and V.~V.~Starshenko,
Mod.\ Phys.\ Lett.\ A {\bf 10}, 235 (1995)
[arXiv:hep-ph/9502348].


\bibitem{COMPASS:2016jwv}
C.~Adolph \textit{et al.} [COMPASS],
Phys. Lett. B \textbf{769}, 34-41 (2017)
[arXiv:1612.00620 [hep-ex]].

\bibitem{E143:1998hbs}
K.~Abe \textit{et al.} [E143],
Phys. Rev. D \textbf{58}, 112003 (1998)
[arXiv:hep-ph/9802357 [hep-ph]].

\bibitem{SpinMuonSMC:1997mkb}
D.~Adams \textit{et al.} [Spin Muon (SMC)],
Phys. Rev. D \textbf{56}, 5330-5358 (1997)
[arXiv:hep-ex/9702005 [hep-ex]].

\bibitem{E155:2000qdr}
P.~L.~Anthony \textit{et al.} [E155],
Phys. Lett. B \textbf{493}, 19-28 (2000)
[arXiv:hep-ph/0007248 [hep-ph]].

\bibitem{RSS:2006tbm}
F.~R.~Wesselmann \textit{et al.} [RSS],
Phys. Rev. Lett. \textbf{98}, 132003 (2007)
[arXiv:nucl-ex/0608003 [nucl-ex]].

\bibitem{HERMES:2006jyl}
A.~Airapetian \textit{et al.} [HERMES],
Phys. Rev. D \textbf{75}, 012007 (2007)
[arXiv:hep-ex/0609039 [hep-ex]].

\bibitem{Deur:2008ej}
A.~Deur, P.~Bosted, V.~Burkert, D.~Crabb, V.~Dharmawardane, G.~E.~Dodge, T.~A.~Forest, K.~A.~Griffioen, S.~E.~Kuhn and R.~Minehart, \textit{et al.}
Phys. Rev. D \textbf{78}, 032001 (2008)
[arXiv:0802.3198 [nucl-ex]].

\bibitem{Deur:2004ti}
A.~Deur, P.~E.~Bosted, V.~Burkert, G.~Cates, J.~P.~Chen, S.~Choi, D.~Crabb, C.~W.~de Jager, R.~De Vita and G.~E.~Dodge, \textit{et al.}
Phys. Rev. Lett. \textbf{93}, 212001 (2004)
[arXiv:hep-ex/0407007 [hep-ex]].

\bibitem{Deur:2014vea}
A.~Deur, Y.~Prok, V.~Burkert, D.~Crabb, F.~X.~Girod, K.~A.~Griffioen, N.~Guler, S.~E.~Kuhn and N.~Kvaltine,
Phys. Rev. D \textbf{90}, no.1, 012009 (2014)
[arXiv:1405.7854 [nucl-ex]].

\bibitem{Deur:2021klh}
A.~Deur, J.~P.~Chen, S.~E.~Kuhn, C.~Peng, M.~Ripani, V.~Sulkosky, K.~Adhikari, M.~Battaglieri, V.~D.~Burkert and G.~D.~Cates, \textit{et al.}
Phys. Lett. B \textbf{825}, 136878 (2022)
[arXiv:2107.08133 [nucl-ex]].

\bibitem{Nachtmann:1973mr}
  O.~Nachtmann,
  Nucl.\ Phys.\ B {\bf 63}, 237 (1973).

\bibitem{Piccione:1997zh}
A.~Piccione and G.~Ridolfi,
Nucl. Phys. B \textbf{513} (1998), 301-316
[arXiv:hep-ph/9707478 [hep-ph]].

\bibitem{Blumlein:1998nv}
J.~Bl\"umlein and A.~Tkabladze,
Nucl. Phys. B \textbf{553} (1999), 427-464.

\bibitem{Burger:2021knd}
S.~B\"urger \textit{et al.} [RQCD],
Phys. Rev. D \textbf{105}, no.5, 054504 (2022)
[arXiv:2111.08306 [hep-lat]].

\bibitem{Sufian:2016hwn}
R.~S.~Sufian, G.~F.~de T\'eramond, S.~J.~Brodsky, A.~Deur and H.~G.~Dosch,
Phys. Rev. D \textbf{95}, no.1, 014011 (2017)
[arXiv:1609.06688 [hep-ph]].

\bibitem{Tanabashi:2018oca}
M.~Tanabashi \textit{et al.} [Particle Data Group],
Phys. Rev. D \textbf{98}, no.3, 030001 (2018)

\bibitem{Herren:2017osy}
F.~Herren and M.~Steinhauser,
Comput. Phys. Commun. \textbf{224}, 333-345 (2018)
[arXiv:1703.03751 [hep-ph]].

\bibitem{Mergell:1995bf}
P.~Mergell, U.~G.~Meissner and D.~Drechsel,
Nucl.\ Phys.\ A {\bf 596}, 367 (1996)
[arXiv:hep-ph/9506375].

\bibitem{Ayala:2018ulm}
C.~Ayala, G.~Cveti\v{c}, A.~V.~Kotikov and B.~G.~Shaikhatdenov,
Eur. Phys. J. C \textbf{78}, no.12, 1002 (2018)
[arXiv:1812.01030 [hep-ph]].

\bibitem{Lepage:1980fj}
G.~P.~Lepage and S.~J.~Brodsky,
Phys. Rev. D \textbf{22}, 2157 (1980).

\bibitem{Pineda:2001zq}
  A.~Pineda,
  JHEP {\bf 0106}, 022 (2001)
  [arXiv:hep-ph/0105008].







\end{thebibliography}
\end{document}